\documentclass[twocolumn,pre,showpacs,amsmath,amssymb,floatfix]{revtex4}

\usepackage{graphicx}
\usepackage{dcolumn}
\usepackage{bm}
\usepackage{amsmath,amssymb}

\begin{document}

\title{Shape and dynamics of seepage erosion in a horizontal granular bed}

\author{Michael Berhanu$^1$}
\altaffiliation[Present address:]{ Mati\`ere et Syst\`emes Complexes (MSC), Universit\'e Paris Diderot, CNRS (UMR 7057), 75013 Paris, France} 
\author{Alexander Petroff$^2$} 
\author{Olivier Devauchelle$^2$} 
\altaffiliation[Present address:]{ Institut de Physique du Globe de Paris-–Sorbonne Paris Cit\'{e}, Universit\'{e} Paris Diderot, UMR CNRS 7154, 1, rue Jussieu, 75238 Paris cedex 05, France}
\author{Arshad Kudrolli$^1$}
\author{Daniel H. Rothman$^2$}
\affiliation{$^1$Department of Physics, Clark University, Worcester, Massachusetts 01610}
\affiliation{$^2$Department of Earth, Atmospheric and Planetary Sciences, Massachusetts Institute of Technology, Cambridge, Massachusetts 02139}

\date{\today}

\begin{abstract}
We investigate erosion patterns observed in a horizontal granular bed resulting from seepage of water motivated by observation of beach rills and channel growth in larger scale landforms. Our experimental apparatus consists of a wide rectangular box filled with glass beads with a narrow opening in one of the side walls from which eroded grains can exit. Quantitative data on the shape of the pattern and erosion dynamics are obtained with a laser-aided topography technique. We show that the spatial distribution of the source of groundwater can significantly impact the shape of observed patterns.  An elongated channel is observed to grow upstream when groundwater is injected at a boundary adjacent to a reservoir held at constant height. An amphitheater (semi-circular) shape is observed when uniform rainfall infiltrates the granular bed to maintain a water table. Bifurcations are observed as the channels grow in response to the ground water. We further find that the channels grow by discrete avalanches as the height of the granular bed is increased above the capillary rise, causing the deeper channels to have rougher fronts. The spatio-temporal distribution of avalanches increase with bed height when partial saturation of the bed leads to cohesion between grains. However, the overall shape of the channels is observed to remain unaffected indicating that seepage erosion is robust to perturbation of the erosion front. 

\end{abstract}

\maketitle

\section{Introduction}
The shape of rivers derives from their complex interactions with landscapes. The flow of water erodes the land which in turn modifies the flow producing meanders and networks of rivers~\cite{izumi1995linear,izumi2000linear,Reitz2010}. Of significant interest is the case of seepage erosion or sapping~\cite{dunne1980formation,higgins1982drainage,bear1979hydraulics}  in which a flow of water occurs inside a permeable and erodible layer of sediment above an impermeable layer of a different composition. Once the water inside this porous medium emerges at the free surface, producing a  spring, the flow removes grains from the surface by erosion, progressively digging a deeper channel, which in turn can draw more water, inducing the growth of a river. Seepage erosion is said to shape many examples of valleys, canyons and river networks and assumed to produce amphitheater-headed valleys~\cite{dunne1980formation,higgins1982drainage,schumm1995ground,abrams2009growth,Petroff},  although recent field reports~\cite{Lamb,lamb2008formation} suggest that some of these examples can be attributed to overland flow as well.  At smaller and more rapid scales, seepage erosion has been considered in the formation of channels on the beach during outgoing tide~\cite{Komar,higgins1982drainage,Otvos1999,Schorghofer}, where capillary cohesion between grains can be also important. The spatial distribution of the groundwater flow when brought from a far away reservoir as opposed to being fed by local precipitation can be crucial to the shape of the channels. 

While the evolution of a river is highly non-linear, the geological variability of the land, initial topography, and changing climate can also make it difficult to identify and extract common features important to their initial formation and growth. Laboratory experiments can help unravel fundamental physical phenomena involved in erosion and the development of  channels. Indeed, seepage erosion, channel formation and bifurcation has been demonstrated with experiments by Howard and collaborators using an inclined granular bed with subsurface water flow injected at a boundary~\cite{howardseepage,howard1988groundwater}, and by a localized and continuous precipitation (mist)~\cite{Gomez}. Further work by our group has shown the effect of slope of the initial bed and the height of the water table on the spacing of channels observed~\cite{Schorghofer}. We have also used the Shields number, which captures the ratio of the viscous drag and gravitational forces, to explain the onset of channelization and slumping of the bed, and its dependence on driving conditions~\cite{Threshold}.  Quantitative topography measurements of the erosion patterns were also used to propose a dynamical model for channelization~\cite{Erosivedynamics}.

Recent experiments by Izumi and collaborators~\cite{Pornprommin2010,Pornprommin2010JGR} have also shown 
channel bifurcation by using coarse sand and a weak slope~\cite{Pornprommin2010}. The authors attribute this occurrence to the important role of the geometry of groundwater flow and the ``resistibility of sediment material to slope failure." Using a similar experimental apparatus, but with overland flow on cohesive soil~\cite{Mezgebu2011}, the same group also observed channel bifurcations. Finally, complex channels produced by a combination of seepage and surface runoff flows have been also investigated~\cite{Ni2006}. 
Thus, it remains unclear if overall channel shape can be used to 
extract if seepage or overland flow is dominant in a given situation and a detailed investigation of both situations is necessary to understand similarities and differences. 

In this paper, we focus on a situation where the granular bed is initially flat and the water table is fed by uniform rain to further connect to examples of seepage channels given in the field, which appear to grow in flat landscapes~\cite{schumm1995ground,abrams2009growth}. To our knowledge, seepage erosion of a flat layer of sand has been never studied experimentally. 
Although this case may appear simpler than the inclined case, it is in fact more complex because the bed slope becomes a free parameter evolving during the growth of the erosion pattern and influencing system dynamics. Further, we study here only the case of erosion driven by seepage, and eliminate  overland flow to simplify the problem. In particular, we seek to understand the influence of the rainfall and the resulting groundwater distribution on the shape and dynamics of resulting erosion patterns. 

The paper is organized as follows. We first introduce the 
essential feature of seepage erosion in Sec.~\ref{physics}, and then we present the experimental apparatus used in our investigation in Sec.~\ref{apparatus}. We then compare in Sec.~IV the obtained erosion patterns when seepage flow is driven by rainfall and by a flow inject at the boundary to simulate groundwater dominated by precipitation outside the box containing the granular bed in the upstream direction.  Then, the effect of depth of the eroded bed is also investigated, enabling us to demonstrate that the erosion driving mechanisms are robust in presence of surface perturbations.  From these measurements, we discuss the influence of rainfall on the shapes of patterns and growth dynamics produced by seepage erosion in Sec.~V. 

\section{Physics of seepage erosion}
\begin{figure}[t]
 \begin{center}
     \leavevmode
    \includegraphics[width=8.6cm]{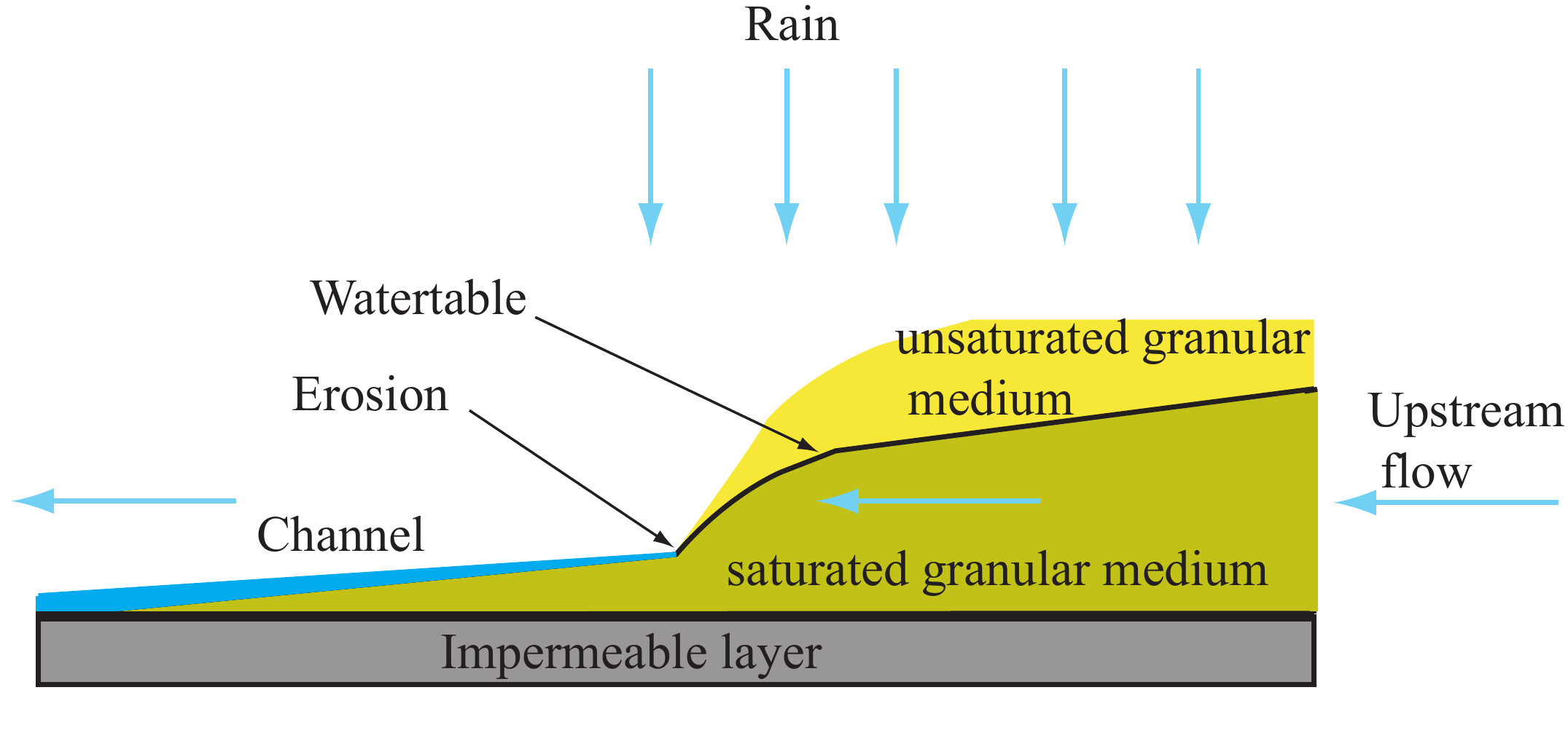}
    \caption{(Color online) A schematic cross-sectional view of the geometry of seepage erosion. Water moves inside the bed from upstream (right) to downstream (left) direction. The water table is the surface separating medium saturated by water from the unsaturated medium, and intersects the granular surface where the water flow emerges at the surface. Grains can be removed by water flowing near the surface leading to headward growth in the upstream direction. 
}
    \label{fig1}
    \end{center}
 \end{figure}

A schematic of a model seepage erosion geometry is shown on Fig.~\ref{fig1}. We consider a horizontal ``step" of porous erodible medium, placed on an impermeable and non-erodible layer. Water moves inside the porous medium fed by uniform rain above with a rate which is lower than the infiltration rate to prevent overland flow. Further, the water can be also fed through the boundary from the upstream direction due to precipitation further upstream or a reservoir. The water table corresponds to the height inside the bed where water saturates the medium and is tilted depending on the flow present, the porosity of the bed, and the boundary conditions. A spring appears where the water table intersect the bed surface, leading to surface flow. The shape of the bank above the surface flow is set by its mechanical stability and conservation of mass. In case of unconsolidated bed composed of sedimentary sand, the slope of the bank is usually given by its angle of repose~\cite{Erosivedynamics}. The surface flow of water erodes the granular bed by removing grains at the bottom of the stream depending on flow rate~\cite{LobkovskyJFM,CharruJFM}, which in turn destabilizes the bank and leads to growth of a channel in the upstream direction. While these overall features are well accepted, a detailed understanding of the  angle at which flow emerges at the surface, its effect on the erosion rate, and development of the erosion front in response to surface perturbations is far from clear.    

\label{physics}
Away from the surface, the flow inside the porous medium is modeled by 
Darcy's law with appropriate boundary conditions. Further, when the vertical component of the flux is small relative to horizontal components, the flow in the porous medium can be approximated as two dimensional, which is called the Dupuit approximation~\cite{bear1979hydraulics}. Therefore the three dimensional problem can be reduced to two dimensions and the seepage flow can be obtained from the water table elevation $h(x,y)$. In the absence of rain, it can be demonstrated that $h^2$ follows the Laplace equation~\cite{bear1979hydraulics}
\begin{equation}
\bigtriangledown^2 h^2 = 0.
\label{laplace}
\end{equation}
with appropriate boundary conditions. In the presence of homogeneous rain, a source term transforms this equation into a Poisson equation
\begin{equation}
\bigtriangledown^2 h^2 = - \dfrac{2 \,P}{k},
\label{poisson}
\end{equation}
where, $P$ is the rainfall rate per unit surface area projected on the horizontal plane, and $k$ the hydraulic conductivity of the porous medium. Because of the important differences in the mathematical properties of these two equations, the shape of the resulting groundwater flow and thus the resulting erosion dynamics are expected to depend strongly on the relative importance of local rainfall compared to upstream flow. 

\begin{figure}[t]
 \begin{center}
     \leavevmode
    \includegraphics[width=8.6cm]{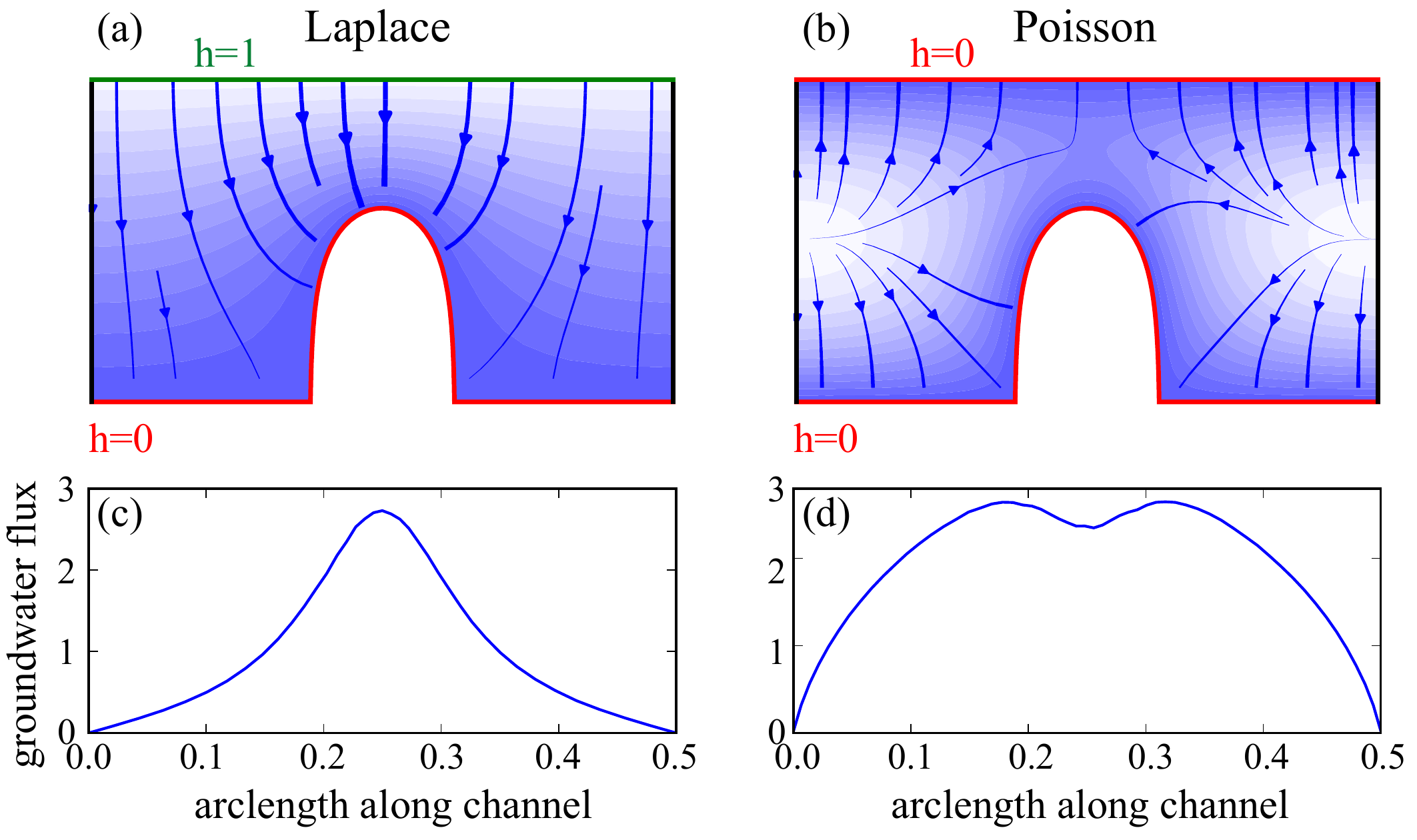}
    \caption{(Color online) (a) Simulation of Eq.~\ref{laplace} in a domain in the presence of a channel. Water is injected on the top boundary ($h=1$) and flows through the bottom boundary with a channel ($h=0$). Contours correspond to $h^2$ and the arrows represent the streamlines of the groundwater flow. (b) Simulation of Eq.~\ref{poisson} in presence of a channel. Water is raining  homogeneously on the domain and flows through the top and the bottom boundaries. (c) Groundwater flux along the arclength of the channel length corresponding to the simulation for the Laplace equation. Maximal flux occurs at the tip of the channel. (d) Groundwater flux along the channel arclength corresponding to the Poisson equation shows two symmetrically placed maxima.}
    \label{fig2}
    \end{center}
 \end{figure}

In order to illustrate the differences in the groundwater flow near a channel under the two different driving conditions, we obtain the shape of the water table by simulating Eq.(1) and Eq.(2) with appropriate boundary conditions [see Fig.~\ref{fig2} (a,b)]. We then find the groundwater flow rate per unit width using Darcy's law: 
\begin{equation} 
Q(x,y) \overrightarrow{n}=- \dfrac{k}{2}\, \overrightarrow{\nabla} h^2 (x,y), \label{Darcy}
\end{equation}
where, $\overrightarrow{n}$ is the unit vector parallel to $\overrightarrow{\nabla} h^2 (x,y)$. We note that in the Laplace case, flow entering the channel is mostly localized near its tip [see Fig.~\ref{fig2}(c)]. Whereas, the groundwater enters the channel almost uniformly from all directions in the Poisson case, and have two symmetrically placed maxima [see Fig.~\ref{fig2}(d)]. Further, we have shown in the past~\cite{Fourremarks} that the local erosion velocity at a channel front increases linearly with the groundwater flow entering at that point. Therefore, when the groundwater flow is driven by an upstream flow (Laplace case), we expect that the channel grows preferentially at its tip. By contrast, water is collected more uniformly on the channel boundary in presence of rain and the maxima of groundwater flux are located on the sides of the channel. Thus, higher erosion rates can be expected near these regions leading to channel splitting. 

Although the thickness of the permeable bed does not affect the seepage flow itself, it does impact the mechanical stability of the bank and the overall channel shape that can develop. In the inclined case, the initial bed slope is the parameter determining the mechanical stability of the sediment~\cite{Schorghofer}. For the flat bed considered here, an analogous parameter may correspond to the ratio between the largest depth of granular material above the water table and the maximum length of the channel and corresponds to the minimum slope inside the channel. Moreover, for small-scale erosion patterns like beach rills, mechanical stability should be strongly affected by capillary cohesion~\cite{Nowak2005}. Above the water table the unsaturated granular medium can become indeed cohesive due to capillary bonds between  particles~\cite{Bocquet2002,Herminghaus2006}, which are present if the sand is wet. By contrast, the saturated granular medium below the water table remains cohesionless. This vertical inhomogeneity of the bed can influence how the material from the bank erodes into the channel~\cite{Fox2007}, before it is removed by the water flow. The occurrence and distribution of these bank collapses determines the overall shape of channel.

\section{Experimental setup}
\label{apparatus}
 \begin{figure}[t]
 \begin{center}
     \leavevmode
  \includegraphics[width=8.6cm]{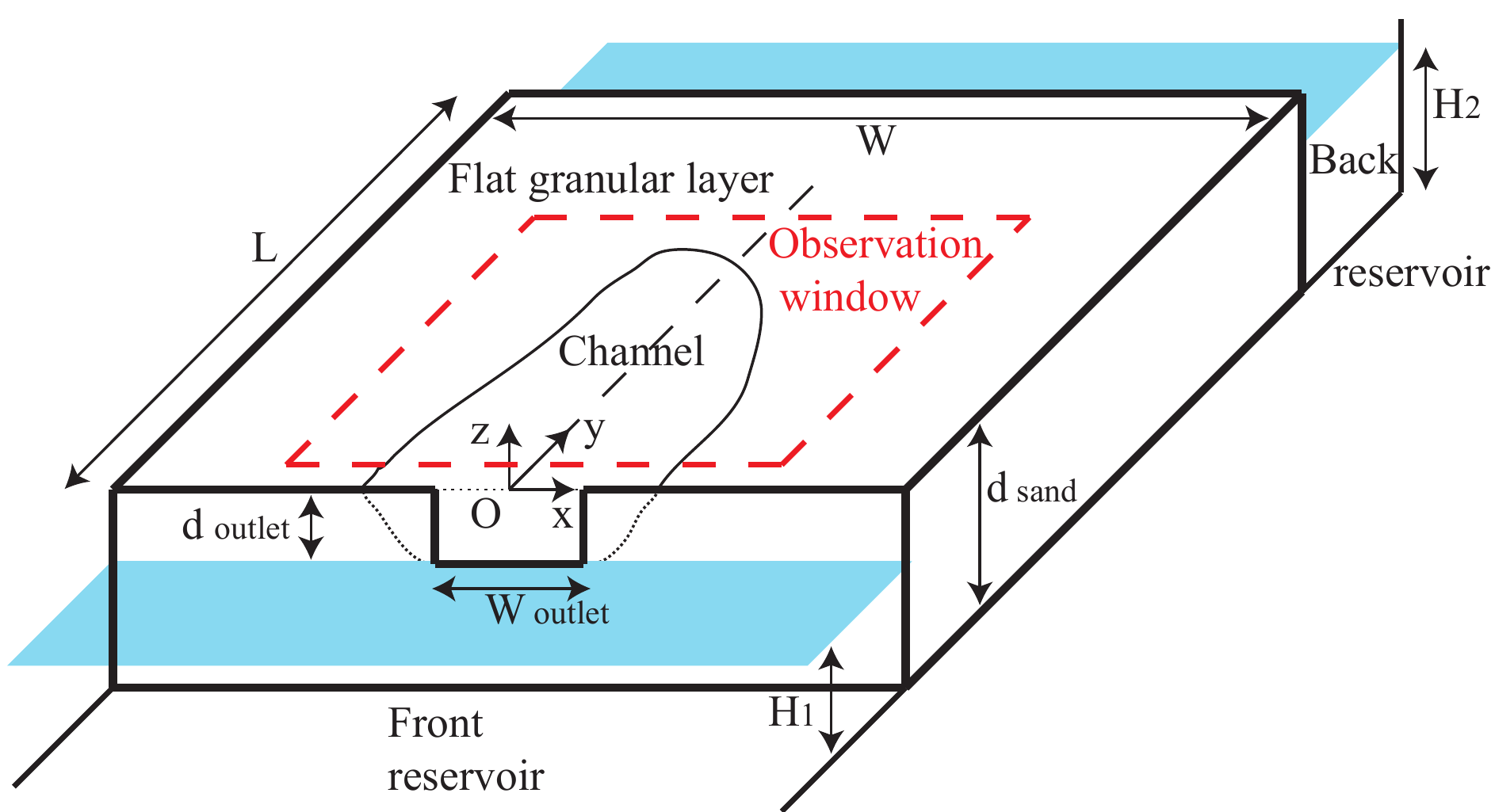}
     \caption{(Color online) Schematic of the experimental apparatus. A Plexiglas box which is open at the top is filled with glass beads with diameter $0.5$\,mm. The front and back side of the granular bed are in contact with water reservoirs maintained at heights $H_1$ and $H_2$ above a boundary which allow water to pass but not the grains. An opening into the front reservoir allows the eroded grains to exit the box. The bed can be exposed to uniform rain produced by an array of three nozzles (not represented). Channel depth corresponds to the vertical elevation difference from the initial flat bed surface height to the bed surface. Topography is measured inside the observation window using a laser scanning technique. Dimensions are given in the text.}
    \label{fig3}
    \end{center}
 \end{figure}

\begin{figure*}
 \begin{center}
    \leavevmode

$\begin{array}{c@{\hspace{.1in}}c@{\hspace{.1in}}c}

\includegraphics[width=7.cm]{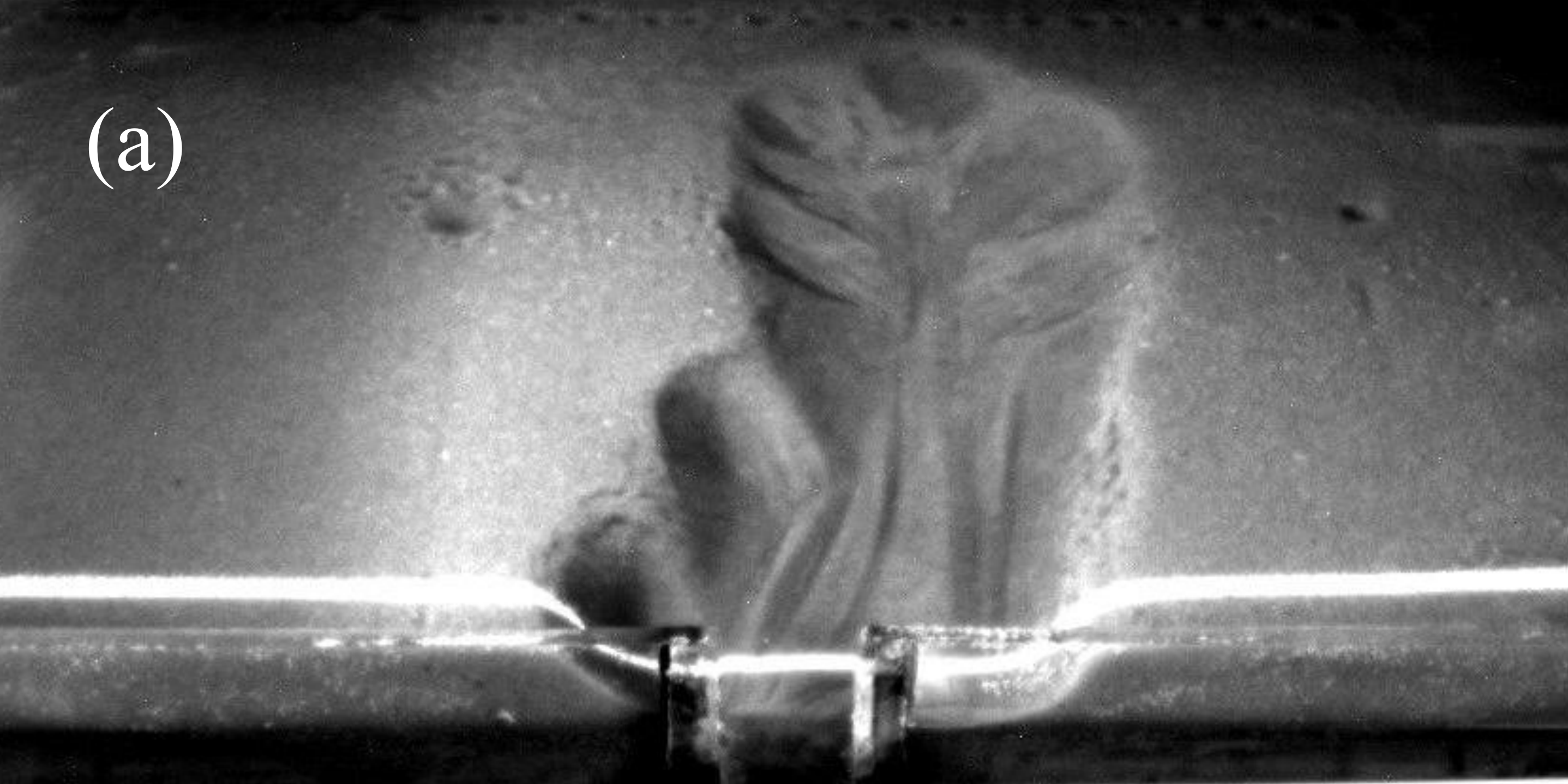}   & \includegraphics[width=7.cm]{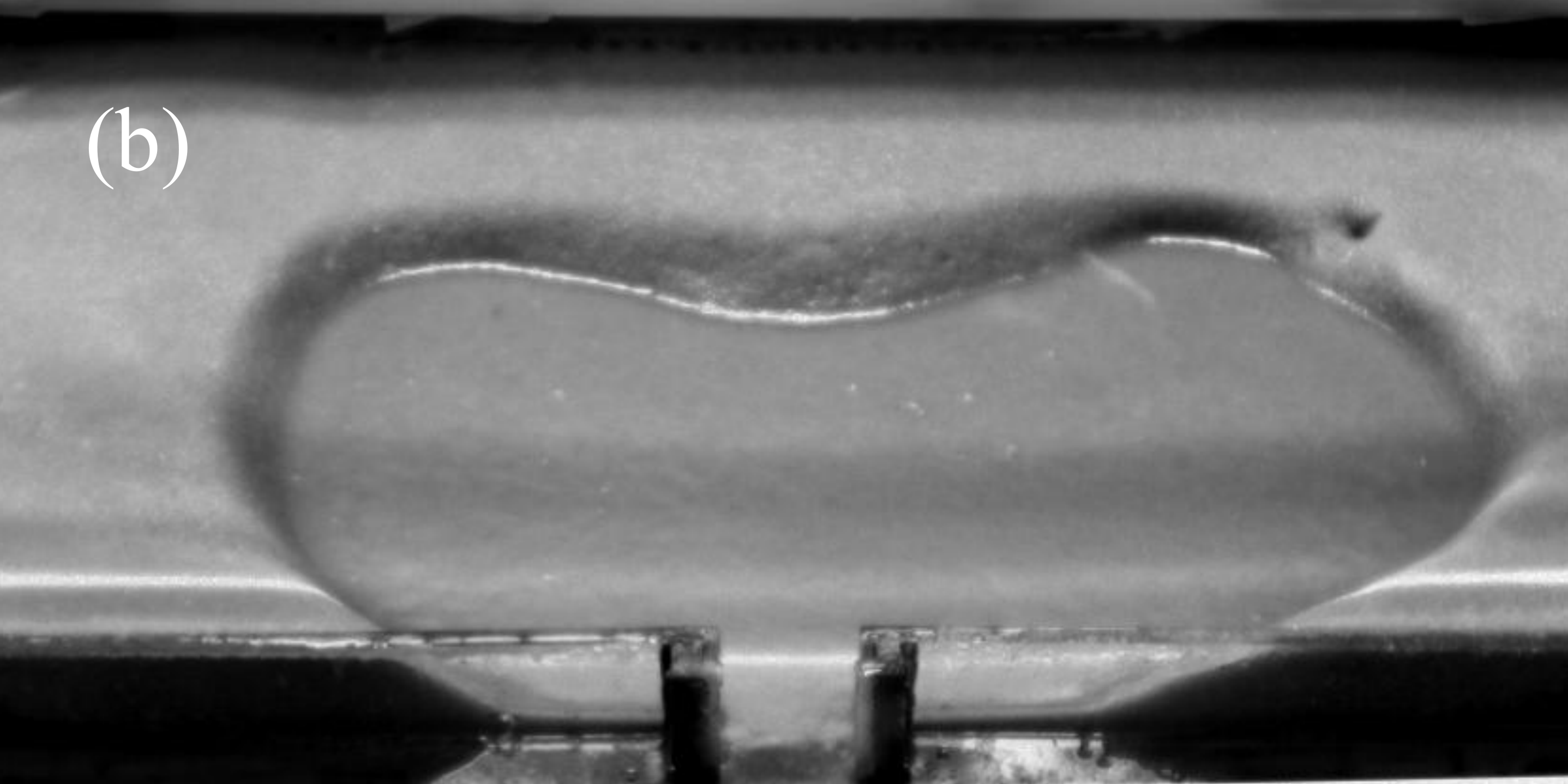} \\
\includegraphics[width=7.cm]{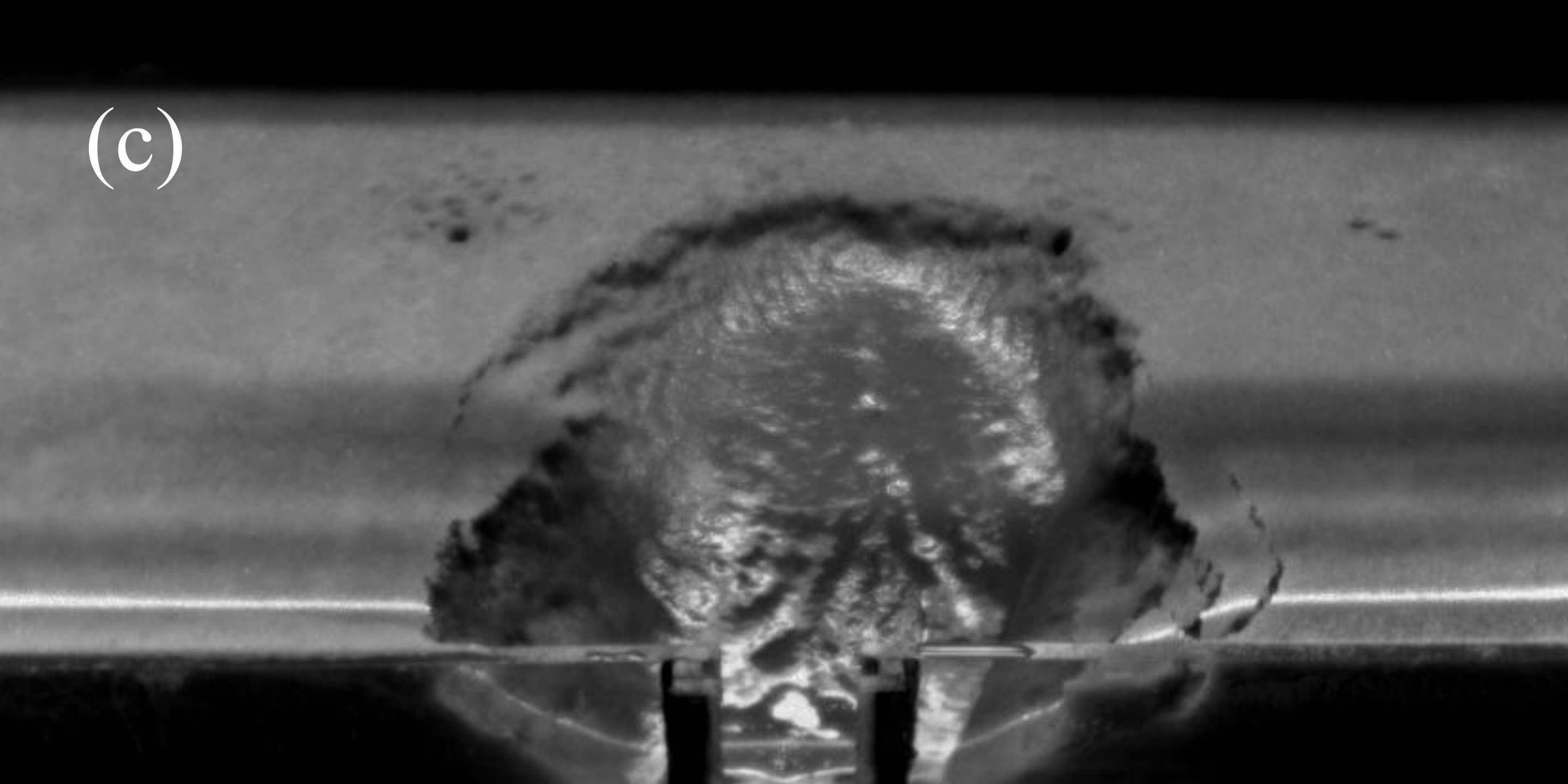}   & \includegraphics[width=7.cm]{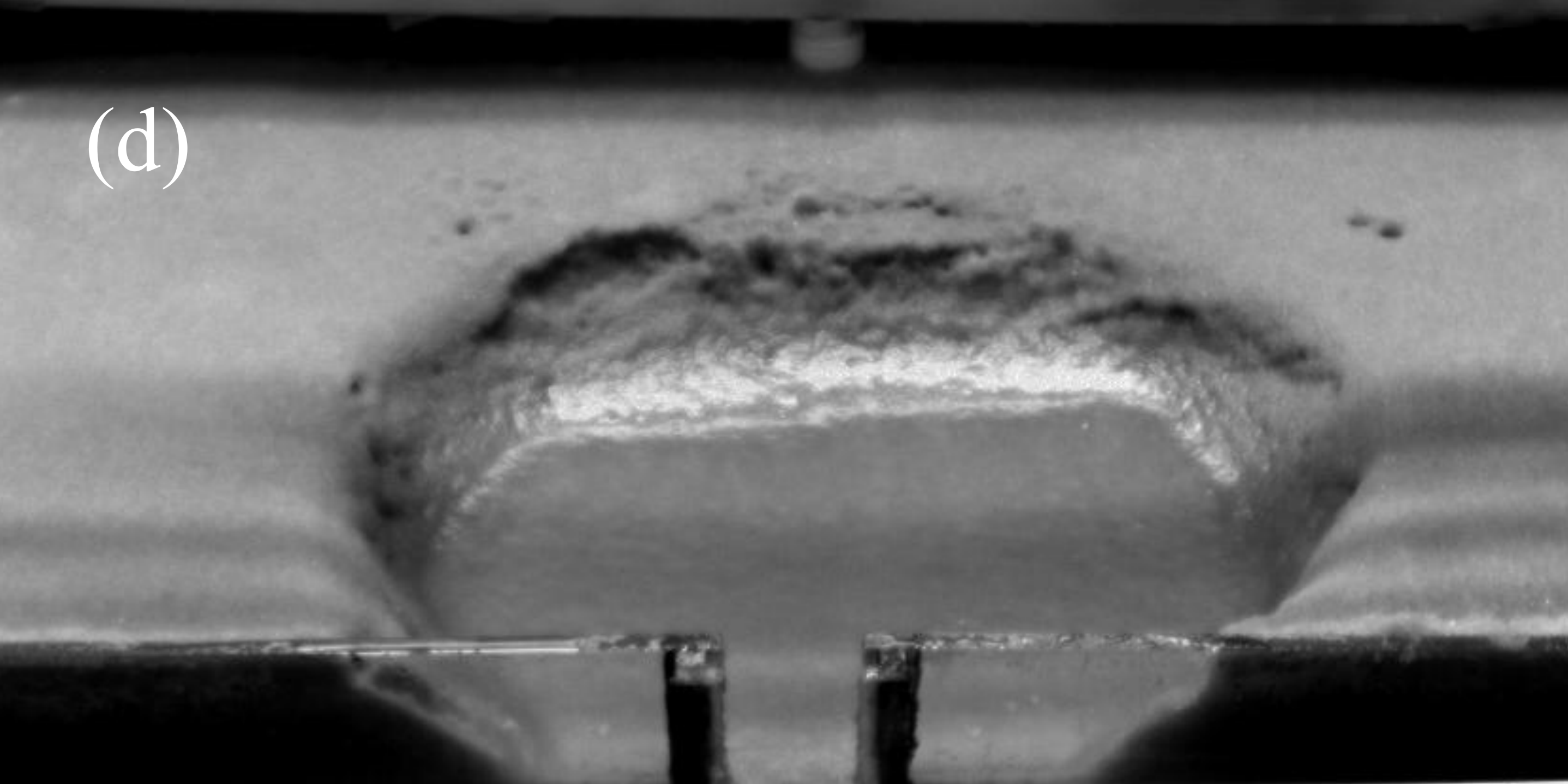} \\
\end{array}$
    \caption{Images of the four kinds of channels observed in the experiments.  
    (a) An elongated channel fed by a flow from a reservoir at the back of the bed after $t=916$\,min with $d_{outlet}=4\,cm$, $\delta_H=5$\,cm and no rainfall ($P=0$). 
    (b) A wider bifurcated  channel fed by rain ($P_1 \sim 1.04 \times 10^{-4}$\,m\,s$^{-1}$) is observed with $d_{outlet}=4\,cm$ ($\delta_H=5$\,cm, $t=595$\,min). (c) An elongated deeper channel with steeper banks is observed with $d_{outlet}=8.0\,cm$, ($P=0$, $\delta_H=6.8$\,cm, $t=311$\,min). (d) The channel is deeper with $d_{outlet}=8.0$\,cm and wider with rain ($P = P_1$,  $\delta_H=2$\,cm, $t=240$\,min).
    }

    \label{Pictures}
    \end{center}
 \end{figure*}
  
The experimental apparatus consists of a Plexiglas box ($120$\,cm wide and $120$\,cm long) which is filled with cohesionless glass beads with diameter $0.5$\, mm as shown in Fig.~\ref{fig3}. The hydraulic conductivity $k$ for this porous material was measured to be $3$\,mm\,s$^{-1}$~\cite{Schorghofer}, and the 
height of capillary rise of water $d_{cap}$ is measured to be $3.2$\,cm. 
Practically, $d_{cap}$  gives a minimum depth of the outlet $d_{outlet}$ to avoid surface flow in the experiments because it gives the height of the water table above where the flow emerges from the granular bed. We label the direction upstream from the outlet as the $y$ axis, its perpendicular in the horizontal as the $x$ axis, and the vertical height as $z$. A metal blade on a track is used to set the initial depth of the bed at $d_{sand}=17.5\,$cm. The bed is also confined (see Fig.~\ref{fig3}) in the $y$ direction between two walls separated by distance $66.5$\,cm, which have a mesh that allow water to pass but not the grains. 
The boundary in the front with the outlet for sand is designed to be permeable to water in order to create a subsurface flow from the back to the front which is unperturbed by the outlet at large distances along the positive and negative $x$ directions. The outlet consists of a $6$\,cm wide rectangular slot as shown in Fig.~\ref{fig3}, with a depth $d_{outlet}$ which can be varied continuously between $11$\,cm and $3.5\,$cm.  Water reservoirs with heights $H_1$ and $H_2$ on the other side of the front and back permeable walls can be adjusted using a water pump to induce a subsurface flow from the back to the front of the bed. 

We also use commercial nozzles which produce small water droplets in a full cone with a $90$\,degree aperture to mimic rainfall on the bed itself. To produce a homogeneous and controlled rainfall rate poses some experimental challenges. After experimentation, we choose to use three nozzles separated by $15$\,cm and placed above the surface at a distance $H_{rain}=46$\,cm off the sand surface. The $y$ coordinate of the central nozzle is $L_{rain}=35$\,cm. We obtained a fairly homogeneous rain in the central region of the bed and within 30\% overall decrease towards the edges. The rain itself can produce a supplementary erosive process of the granular bed. For example, rainfall on a cohesive material can create complex landscapes~\cite{Bonnet2009}. In our experiments, the impact of the water drops act to only smoothen spatial variations of the granular surface and does not modify the seepage flow~\cite{Ellison1950}. Experiments with rain are conducted with a total flow rate of 69\,cm$^3$ s$^{-1}$ with 10\% fluctuations. The corresponding rainfall rate $P_1 \sim 1.04 \times 10^{-4}$\,m\,s$^{-1}$ is obtained as the ratio of rain per unit time and the surface area in a central zone. To avoid any excess water at the boundaries of the box, a gutter is also placed right next to it to remove water which hits the side boundary. Finally due to the presence of this gutter and the blade system, the actual length in the $y$ direction over which experiments are conducted is $L=55$\,cm. 

A number of parameters can be varied in our experiments. These include the size of the outlet $d_{outlet}$, which controls the eroded depth by setting its maximal possible value, the amplitude of the upstream flow by setting the difference between $H_2$ and $H_1$, $\delta_H = H_2 - H_1$. In order to simplify the experimental situation, we set $H_1$ to the same level as the bottom of the outlet i.e. $H_1 = d_{sand} - d_{outlet}$. Further, presence or absence of rainfall is simulated by turning on and off  water circulation to the nozzles. 
The amplitude of the upstream flow rate due to $\delta_H$ is proportional in the framework of Dupuit approximation with the parameter $U=\dfrac{{H_2}^2-{H_1}^2}{L^2}$~\cite{bear1979hydraulics}, where $L$ is the length of the bed. This corresponds to a dimensionless seepage velocity due to the upstream flow alone. We note that for the Dupuit approximation to be valid, we need $U \ll 1$. Finally the dimensionless rainfall rate is $\dfrac{2 {P}}{k}$ and corresponds to $0$ or $0.069$ depending on the absence and presence of local rain in our experiment.

We try to estimate the flow rate at the position of the outlet $(x=0, y=0)$, using the Darcy law, Eq.~\ref{Darcy} . With the approximation that the seepage flow depends mainly on $y$, we obtain:
$$Q_w=\dfrac{k}{2}\, \left( \dfrac{{H_2}^2-{H_1}^2}{L^2} \right)\, L$$
with a difference in height of the front and back reservoir and without uniform rain. With only uniform rain, we obtain: 
$$Q_r = P\, L,$$
where, $L$ is the length of the bed in the $y$ direction, and $W$ is the total width of the sand box. For a significant height difference $H_1=9.5$\,cm and $H_2=16.3$\,cm, the flow rate can be estimated to be $Q_w \sim 5\,\times\,10^{-6}\,$m$^2$s$^{-1}$. Whereas the flow rate due to rain without height difference is $Q_r \sim 1\times\,10^{-4}\,$m$^2$s$^{-1}$.
Therefore, the seepage flow rate is at least one order of magnitude larger due to rain and thus dominant in our experiment.

The surface is observed with a CMOS camera located approximately $2$\,m above the granular bed.  For quantitative measurements of channel shape, we use a laser scanning technique~\cite{Erosivedynamics} to reconstruct the topography inside an observation window (-40\,cm $ < x < 40$\,cm  and 5\,cm $< y < 40$\,cm) indicated in Fig.~\ref{fig3}. A laser sheet is swept over the bed surface along the $y$ axis and the illuminated crossection of the bed at that location is imaged and used to reconstruct the topographic map of the bed after appropriate calibration. The time when erosion is observed to start is chosen as time $t=0$, and is studied over several hours with successive scans. 
Experimental analysis is terminated when channel size exceeds the size of observation window, which also avoids direct effect of the boundaries on the growth of the channels. We also confirmed that the presence of water flow at the surface few millimeter deep does not affect the topography measurements.

\begin{figure*}[!t]
 \begin{center}
\includegraphics[width=8.9cm]{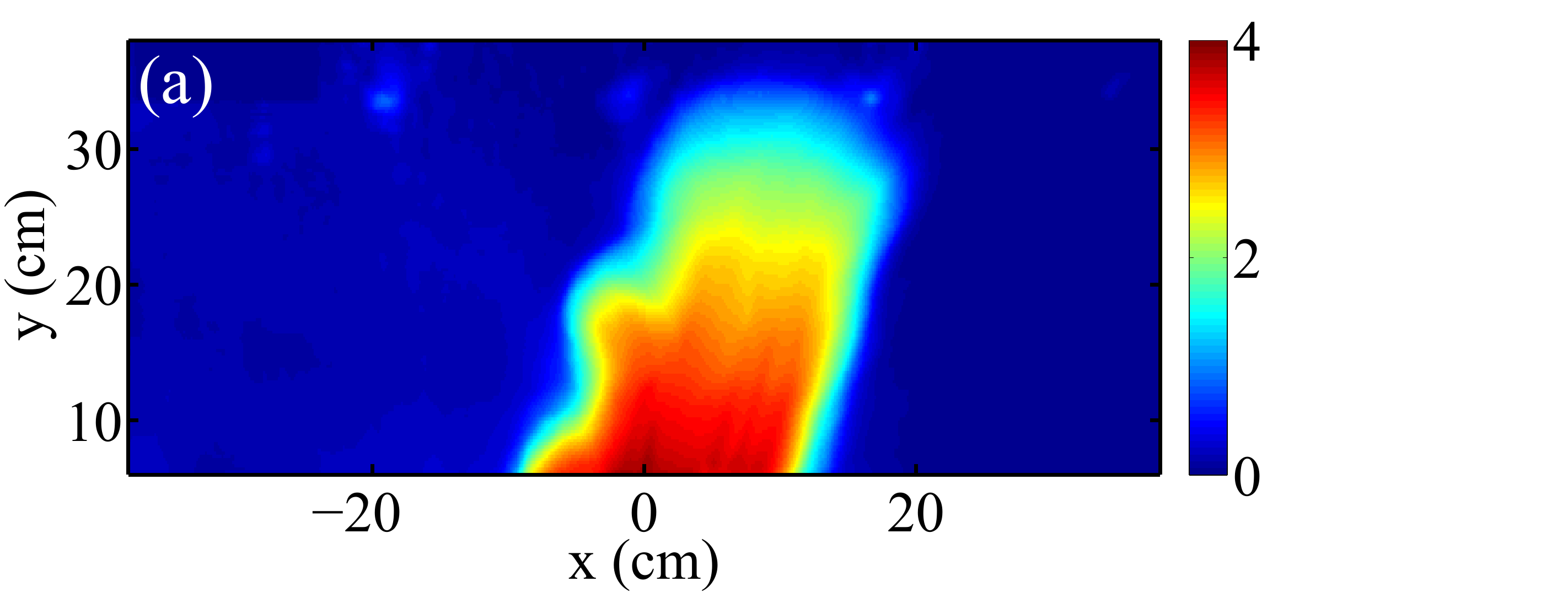}  
  \includegraphics[width=8.9cm]{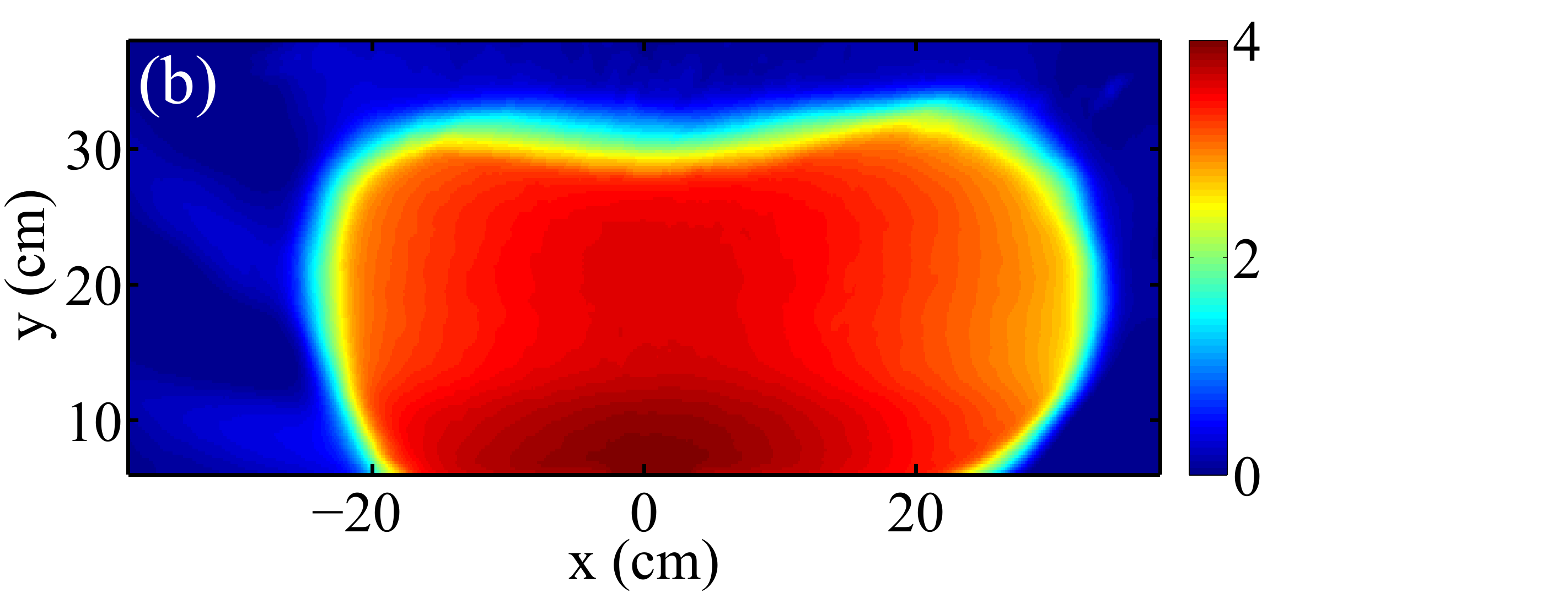} \\
\includegraphics[width=8.9cm]{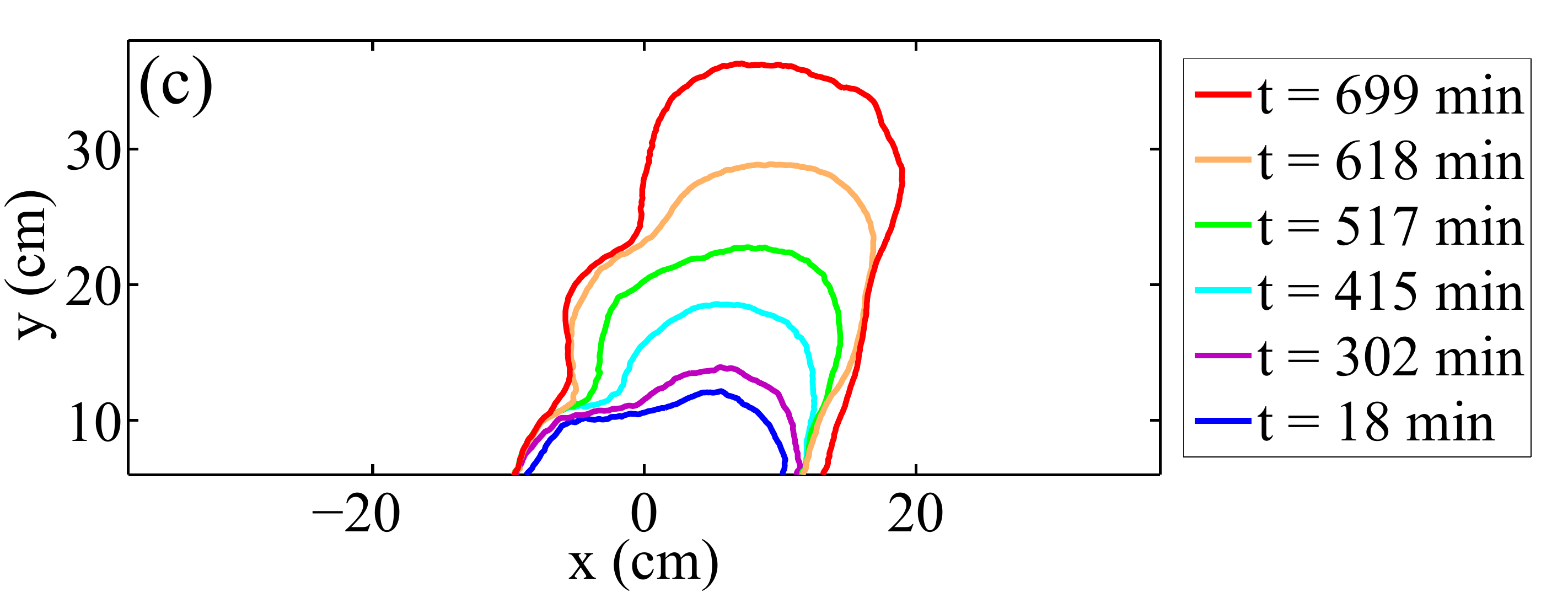} 
\includegraphics[width=8.9cm]{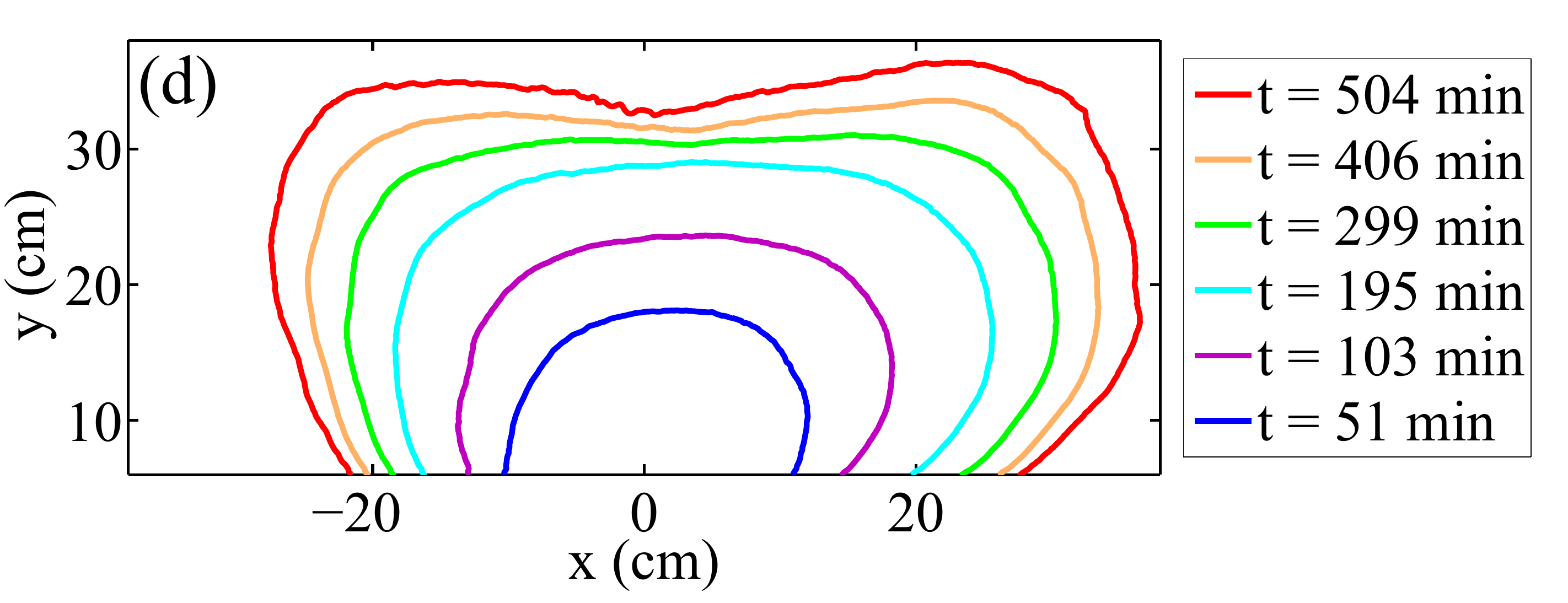} \\
    \caption{(Color online) Comparison of channels in a shallow channel regime where the capillary rise is of order of the channel depth. (a) Depth map of the channel observed after $t=677$\,min with $\delta_H=5$\,cm, $P=0$ and $d_{outlet}=4$\,cm. Depth is indicated with a scale in centimeters measured from the initial flat bed. (b) Depth map of channel observed after $t=300$\,min with  $P_1 \sim 1.04 \times 10^{-4}$\,m\,s$^{-1}$, $\delta_H=0.9$\,cm and $d_{outlet}=4$\,cm. (c, d) The corresponding channel shape evolution for the depth  $d(x,y)=1$\,cm contour increase in length with time.}
    \label{fig5}
    \end{center}
 \end{figure*} 
 
 \begin{figure}[!h]
 \begin{center}
\includegraphics[width=4.25cm]{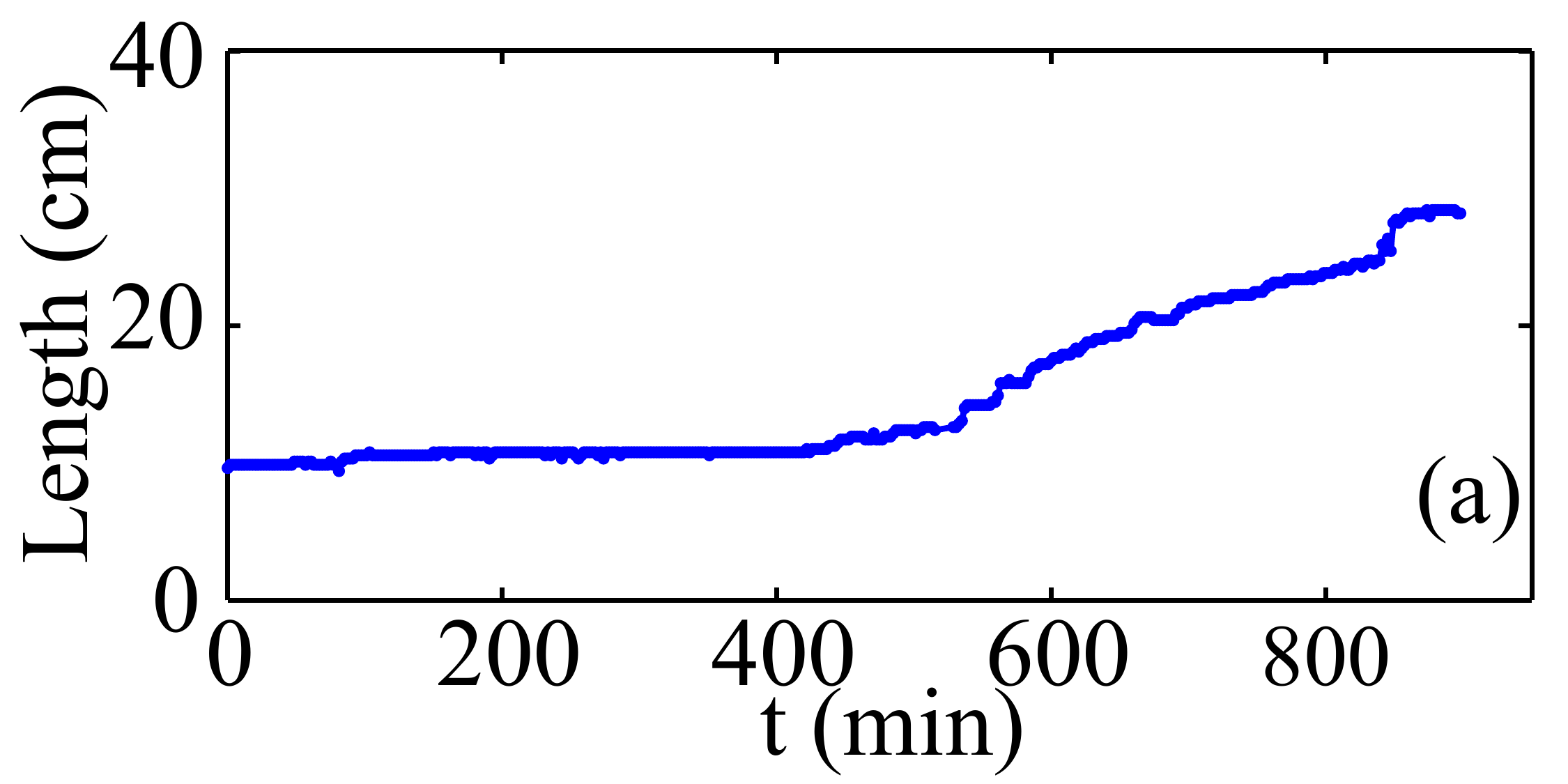} 
\includegraphics[width=4.25cm]{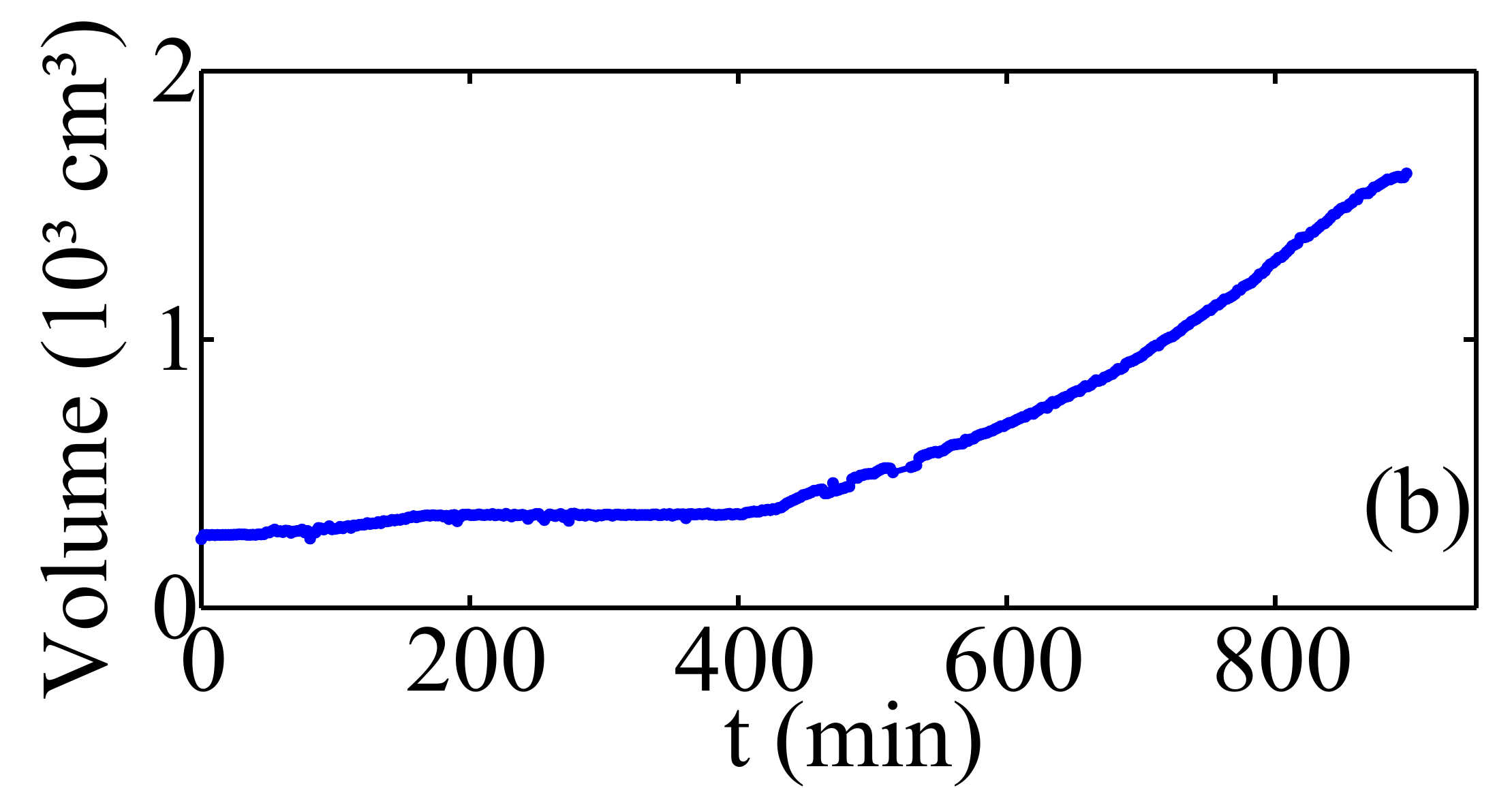}\\
\includegraphics[width=4.25cm]{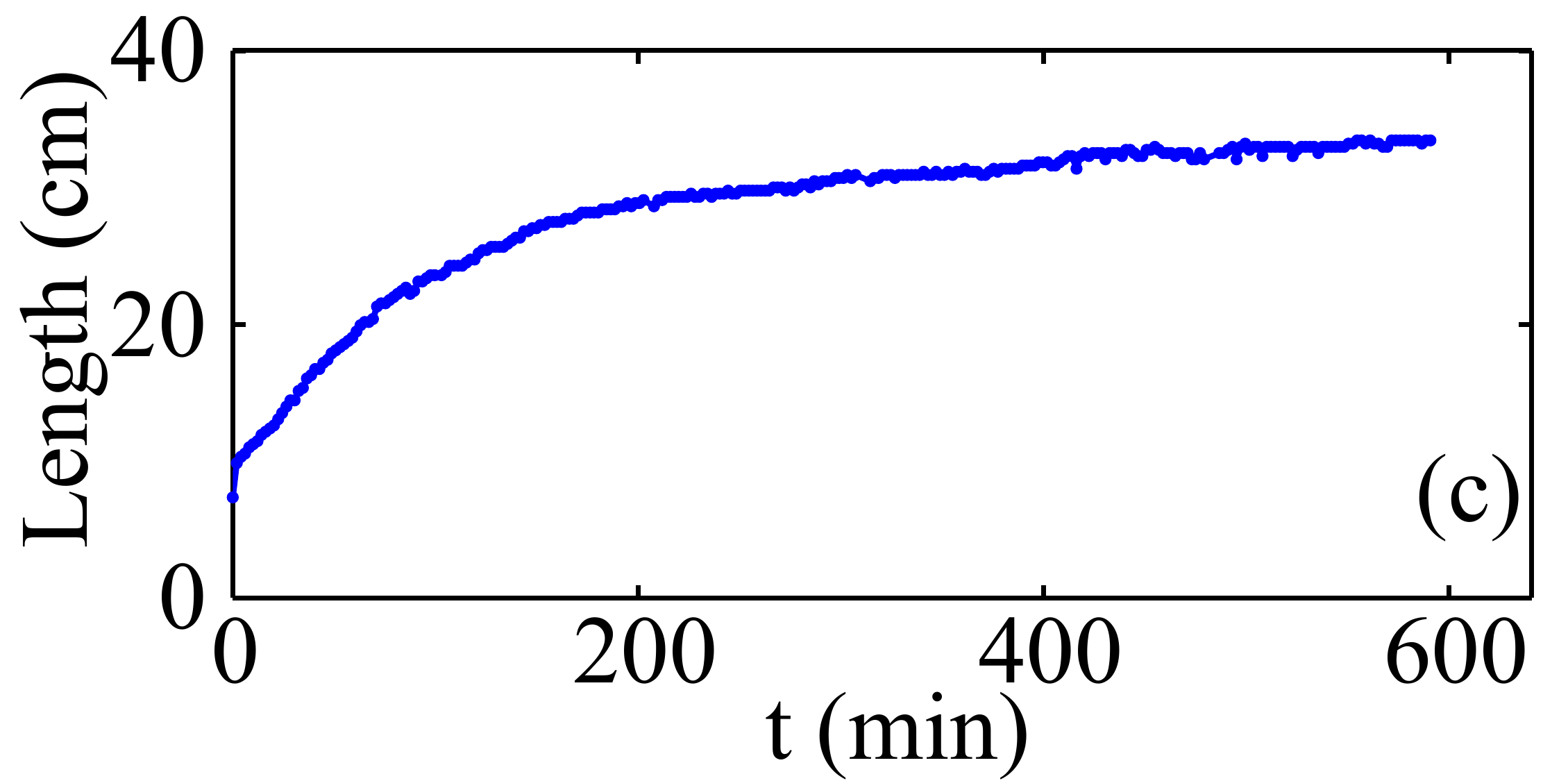} 
\includegraphics[width=4.25cm]{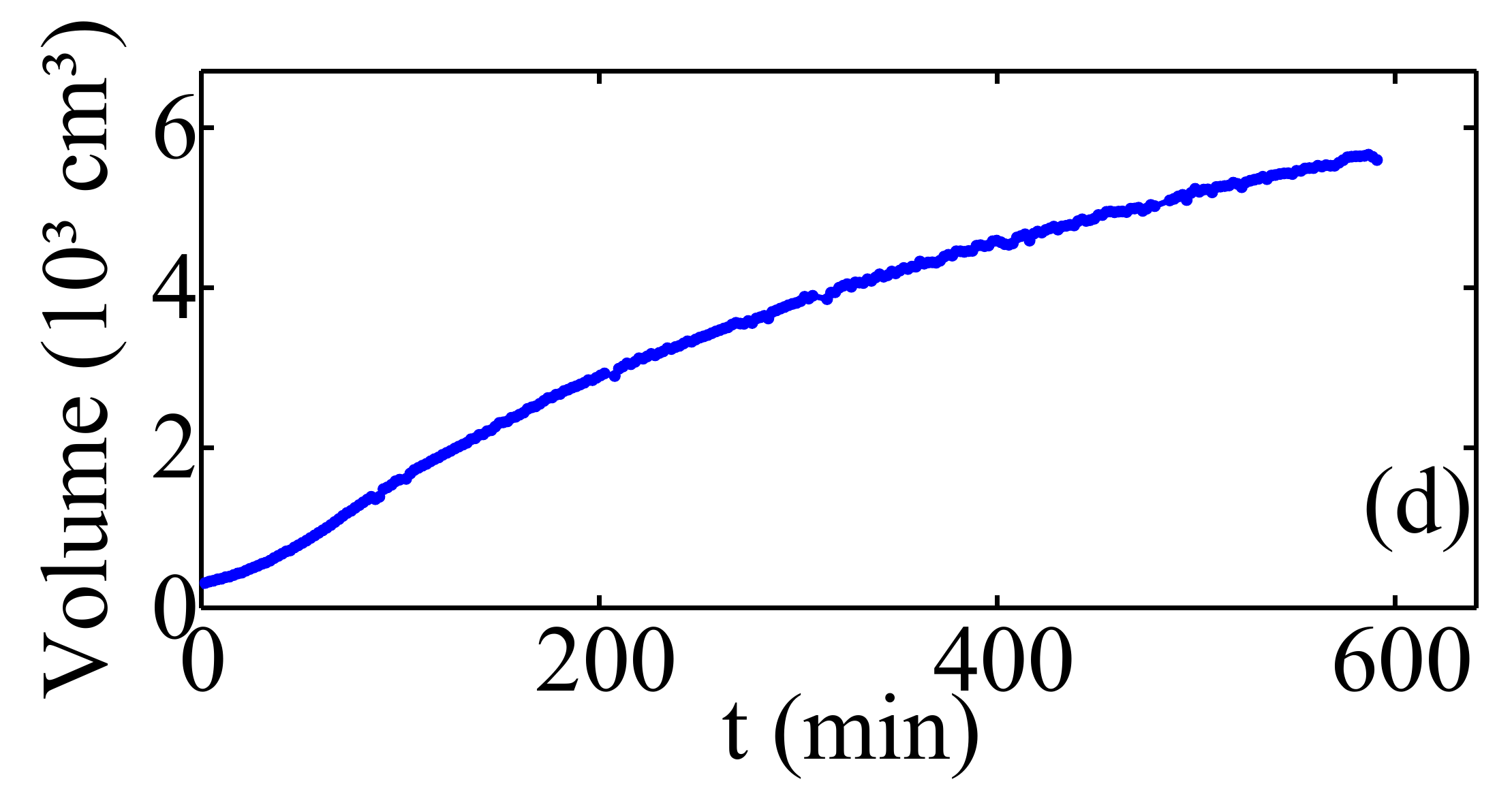} \\
        \caption{(Color online) (a,b) Channel length and eroded volume evolution in absence of rain  $P=0$  with $\delta_H=5$\,cm. 
        (c,d) Channel length and eroded volume evolution in presence of rain $P = P_1$ with $\delta_H=0.9$\,cm. 
         }
    \label{fig6}
    \end{center}
  \end{figure} 

\section{Observations}
For a sufficiently high seepage rate, erosion is observed near the outlet with a sediment flux rate which is related to the seepage flow rate. The erosion front can stop after progressing in the upstream direction or continue till the boundary is reached depending on the applied conditions. We call the resulting patterns channels because they are formed as a result of erosion and a stream is present at the bottom of the eroded bed, even though the ratio of the length and the width may be comparable and not as high as in case of channels in the conventional sense. The grains removed through the outlet do not play a further role in the erosion back stream.  For sufficiently large  $\delta_H$, or in cases where $d_{outlet} < d_{cap}$, the entire granular bed becomes saturated and a surface flow occurs over the bed which is different from seepage erosion~\cite{izumi1995linear,izumi2000linear}, and therefore not discussed further here.

Because of the bed geometry and the difficulty in actually measuring the water flow on and near the surface, it is not possible to compute the relevant Shields number. We notice nevertheless that in our experiments, surface flows cover a large area of channels with a depth of a few millimeters which is at least one order of magnitude smaller than the channel depth. Consequently, we focus our study on shape evolution and growth of channels as a function of the type of seepage flow and as a function of channel depth controlled by the height of the outlet $d_{outlet}$. A variety of channel shapes and growth velocities can be obtained using the reservoir or uniform rainfall and various outlet depths. Figure~\ref{Pictures} shows images of four different examples obtained by using various combination of outlet depth and sources of seepage flow. 

\subsection{Effect of the spatial distribution of seepage water source on channelization}
We begin by discussing the examples for small values of $d_{outlet}$ where the Dupuit approximation is valid, and where the channel depth is just above the threshold needed to prevent overland flow. 
The map of a resulting channel depth $d(x,y)$ relative to the initial surface obtained with laser aided topography after time $t = 677$\,min is shown in Fig.~\ref{fig5}(a) for  $d_{outlet}=4.0$\,cm, $\delta_H=5$\,cm, and $P=0$. A smooth finger shaped channel is obtained with more or less uniform internal slope. To illustrate the evolution of the shape, we plot the contour $d(x,y)=1$\,cm corresponding to the external boundary of the channel at various times in Fig.~\ref{fig5}(c). The growth appears directed to the back of the bed and the channel width remains constant. By contrast a significantly wider channel is obtained when groundwater is fed by uniform rain (see Fig.~\ref{fig5}b). The evolution of the corresponding $d(x,y)=1$\,cm contour is shown in Fig.~\ref{fig5}(d). We observe that the growth is isotropic initially, but appears to then grow towards the top left and right corners as the channel starts to grow after a distance of about $15$\,cm from the outlet. 

As illustrated in Fig.~\ref{fig2}(b,d), we expect that the groundwater flux entering into the channel has two maxima on each channel sides in the presence of homogeneous rain. As erosion rate should be higher at these points, a channel splitting could be so induced. This bifurcation could be the first step toward creation of a river network, such as the seepage networks of the Florida Panhandle~\cite{abrams2009growth,Petroff}. However, the typical channel width (which is set by the flow rate and the size of grains)~\cite{Longitudinal} is large compared to the system size to observe further channel splitting. By repeating experiments under similar conditions, we find that the bifurcation always occurs but in general not symmetrically and can sometimes take more complex shapes. This variability may indicate well an instability in addition to a bifurcation driven by groundwater flow. Finally, the small scale topography inside the channel visible on the pictures Fig.~\ref{Pictures} (a-b), appears more complex in the case where $P=0$. The surface flow is greater when $P = P_1$, further eroding the channel bottom, making it nearly flat in our experiments. Falling droplets also appear to smooth small surface perturbations~\cite{Ellison1950} and the channel bank appears more regular with rain.

\begin{figure*}[t]
 \begin{center}
\includegraphics[width=8.9cm]{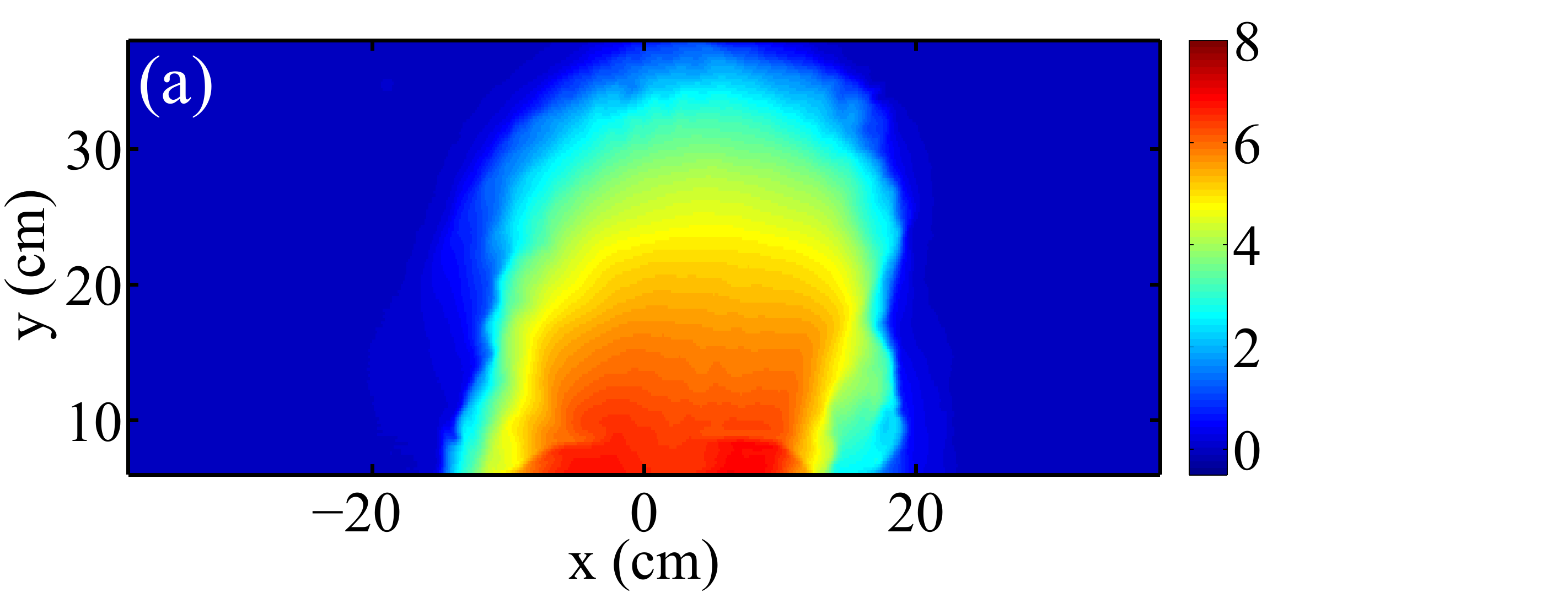}  
\includegraphics[width=8.9cm]{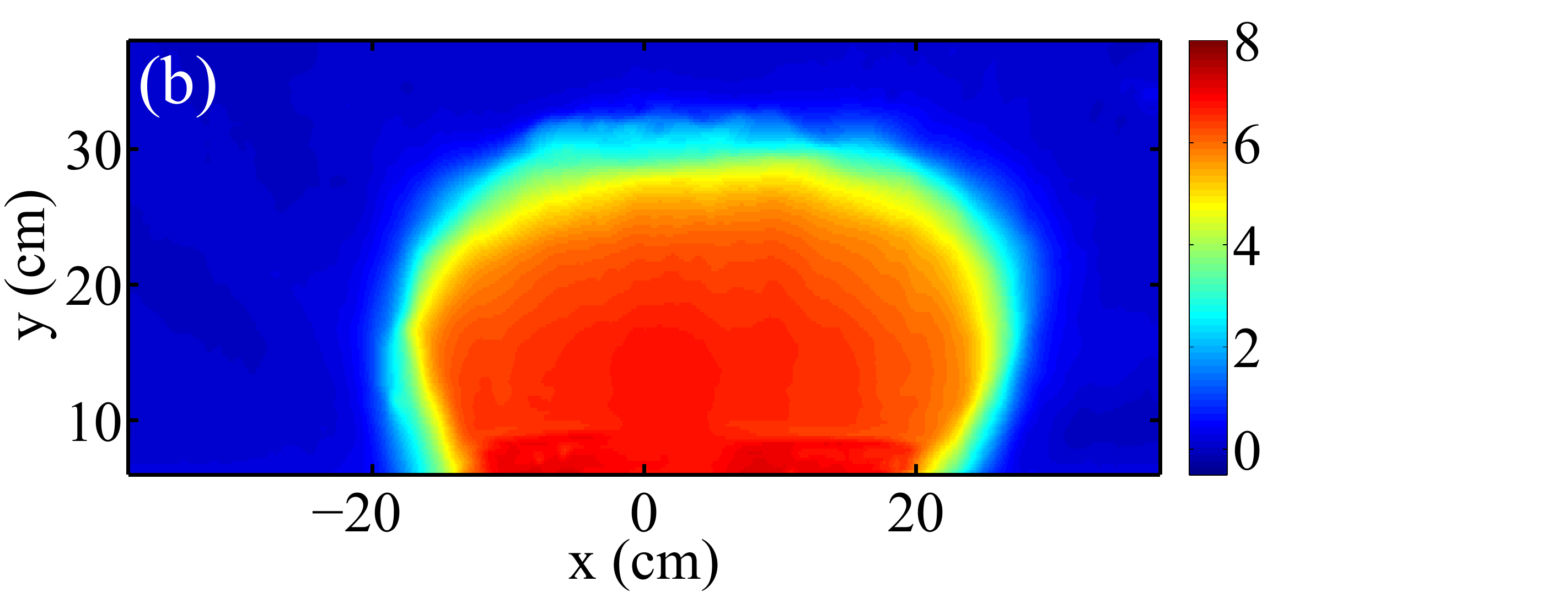} \\
\includegraphics[width=8.9cm]{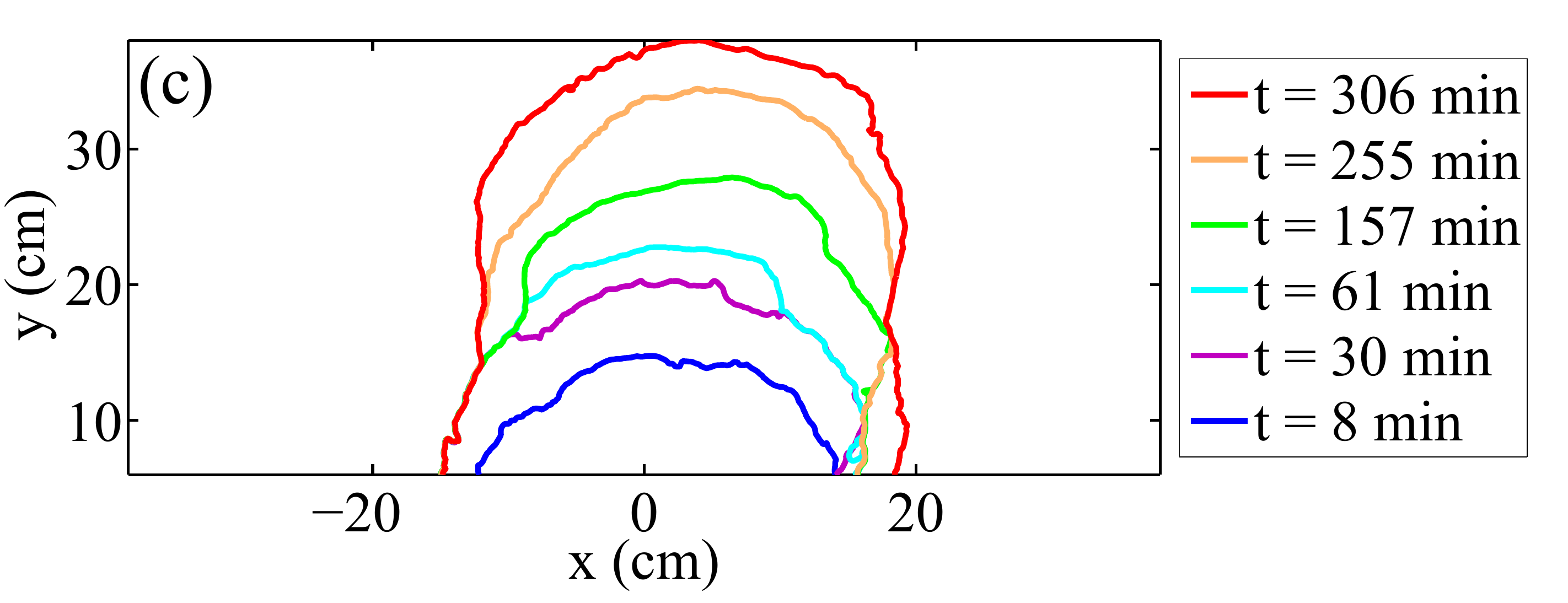} 
\includegraphics[width=8.9cm]{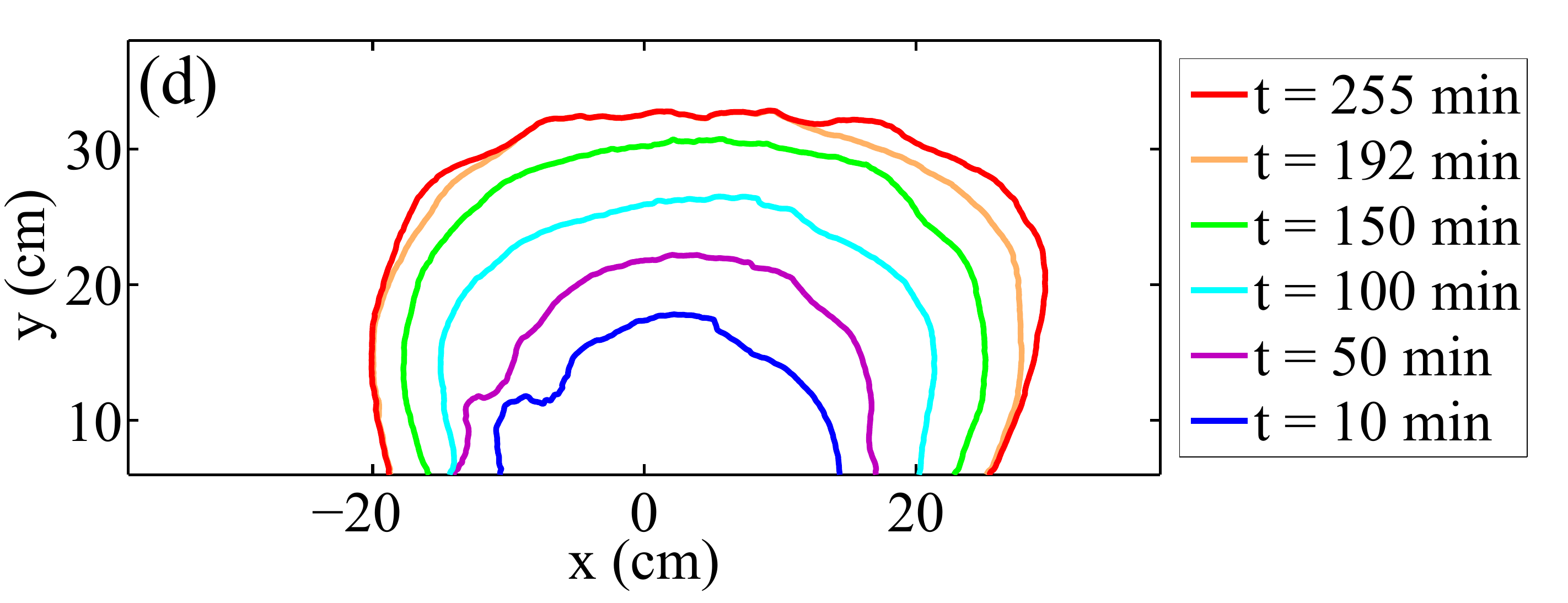} \\
    \caption{(Color online) Comparison of channel shape in the regime where $d_{outlet} > d_{cap}$ and significant layer of unsaturated cohesive bed layer is present. 
    (a) Depth map of channel observed after $t=306$\,min with $d_{outlet}=8.0$\,cm, $\delta_H=6.8$\,cm and $P=0$. (b) Depth map of channel observed after $t=225$\,min with $d_{outlet}=8.0$\,cm, $\delta_H=2.0$\,cm and $P = P_1$. (c, d) The corresponding channel shape evolution for the depth  $d(x,y)=1$\,cm contour. The contours increase in length with time.}
    \label{fig7}
    \end{center}
 \end{figure*} 
 
\begin{figure}
 \begin{center}
\includegraphics[width=4.25cm]{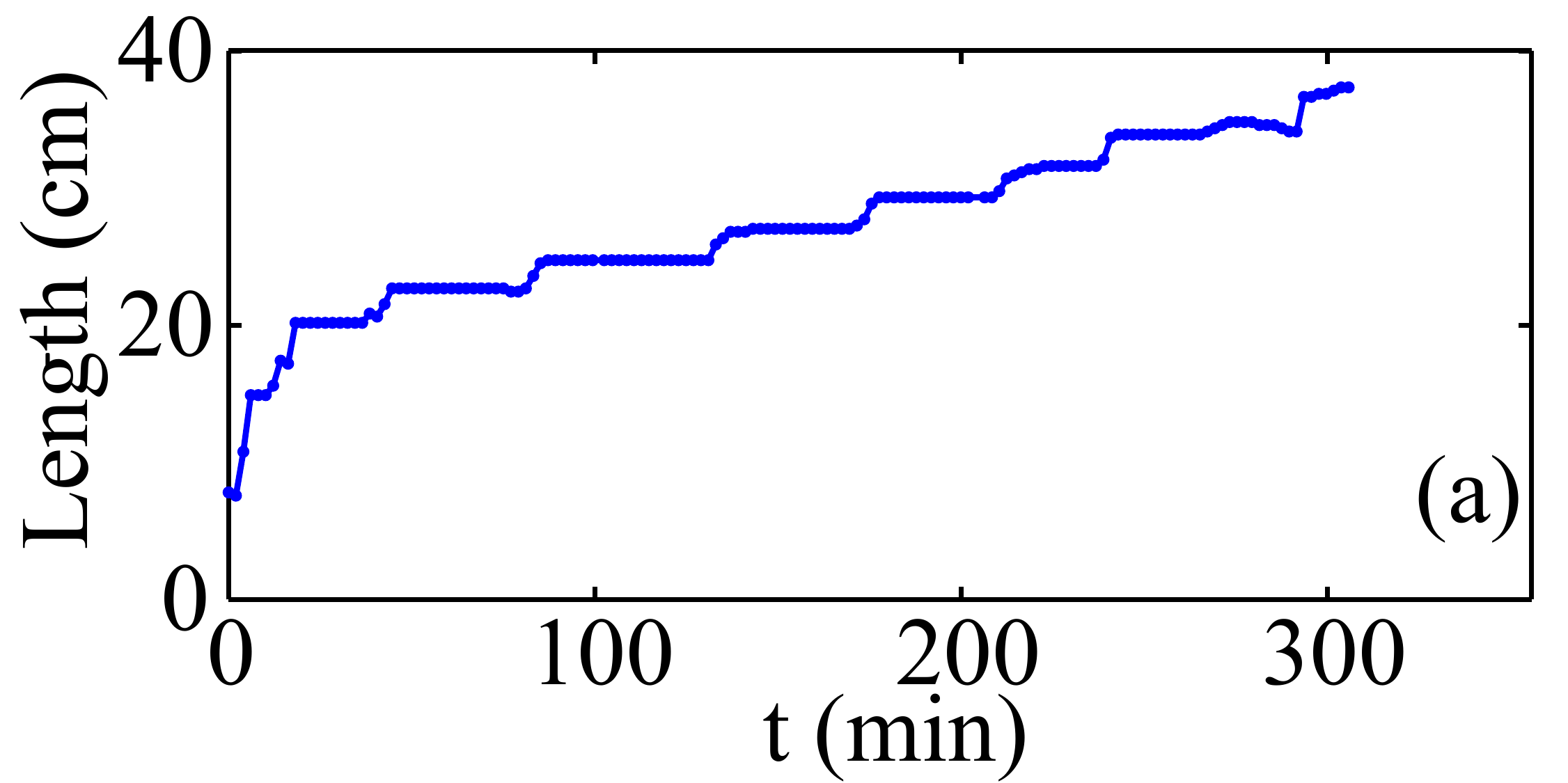} 
\includegraphics[width=4.25cm]{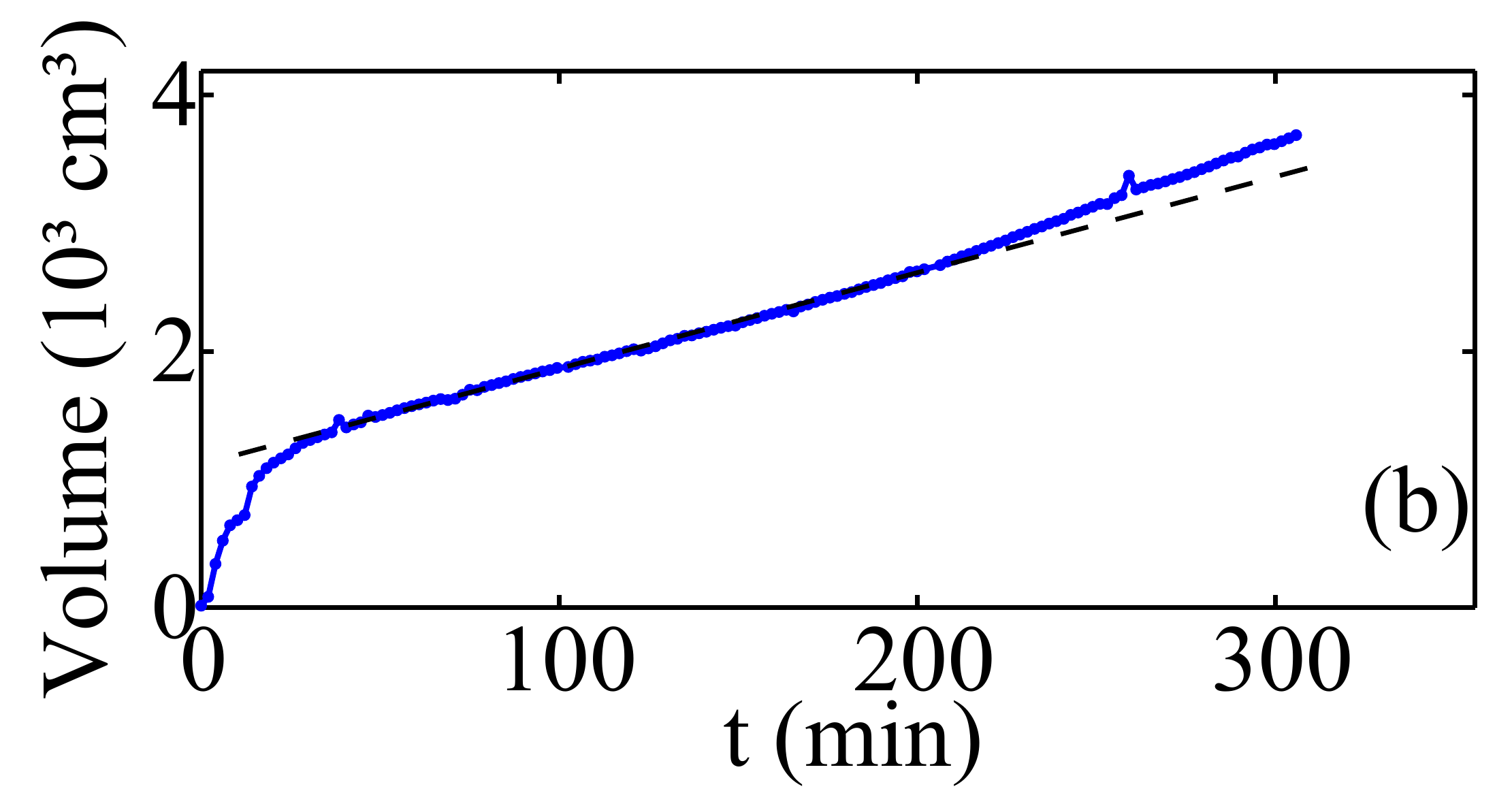} \\
\includegraphics[width=4.25cm]{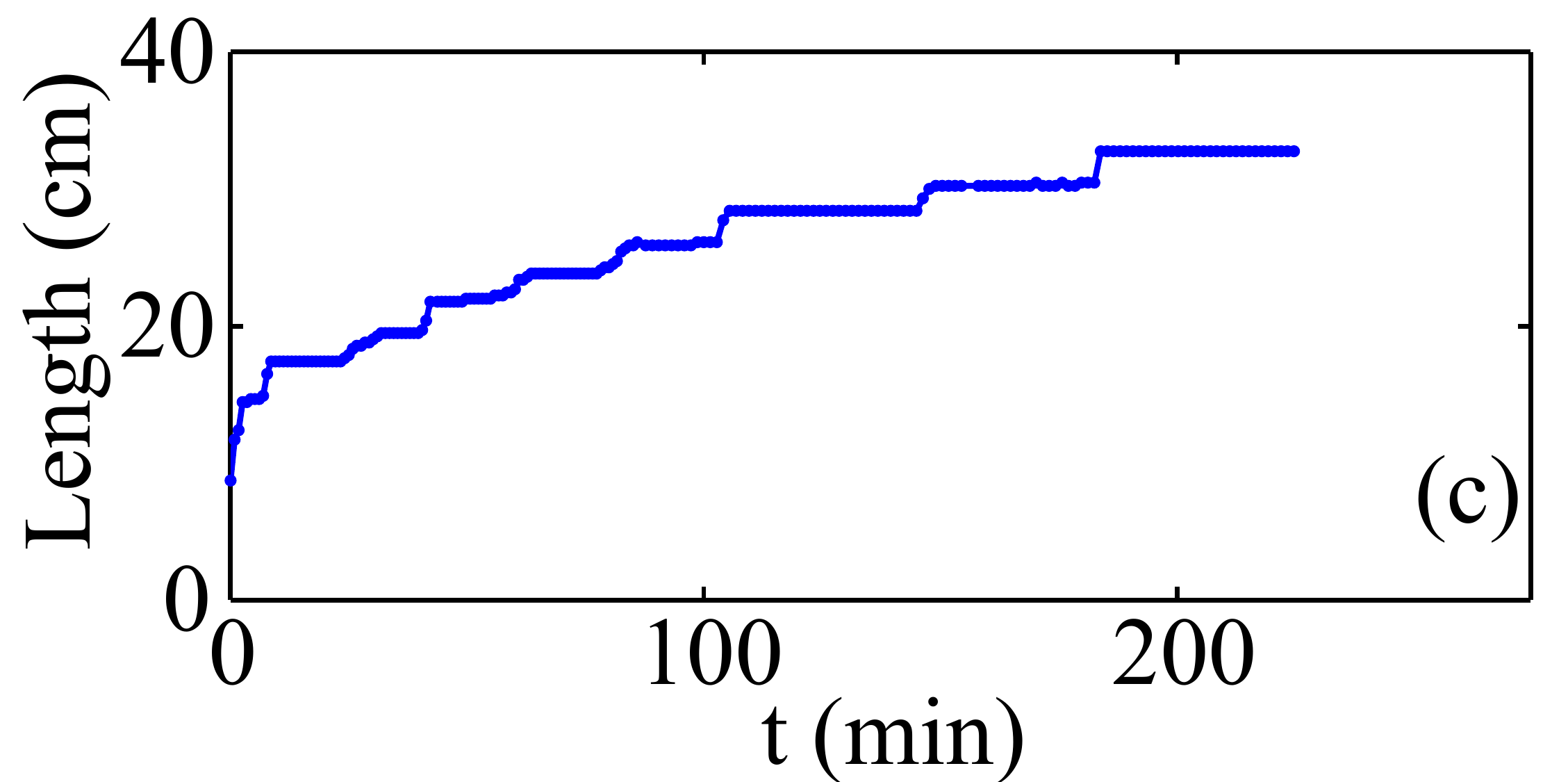} 
\includegraphics[width=4.25cm]{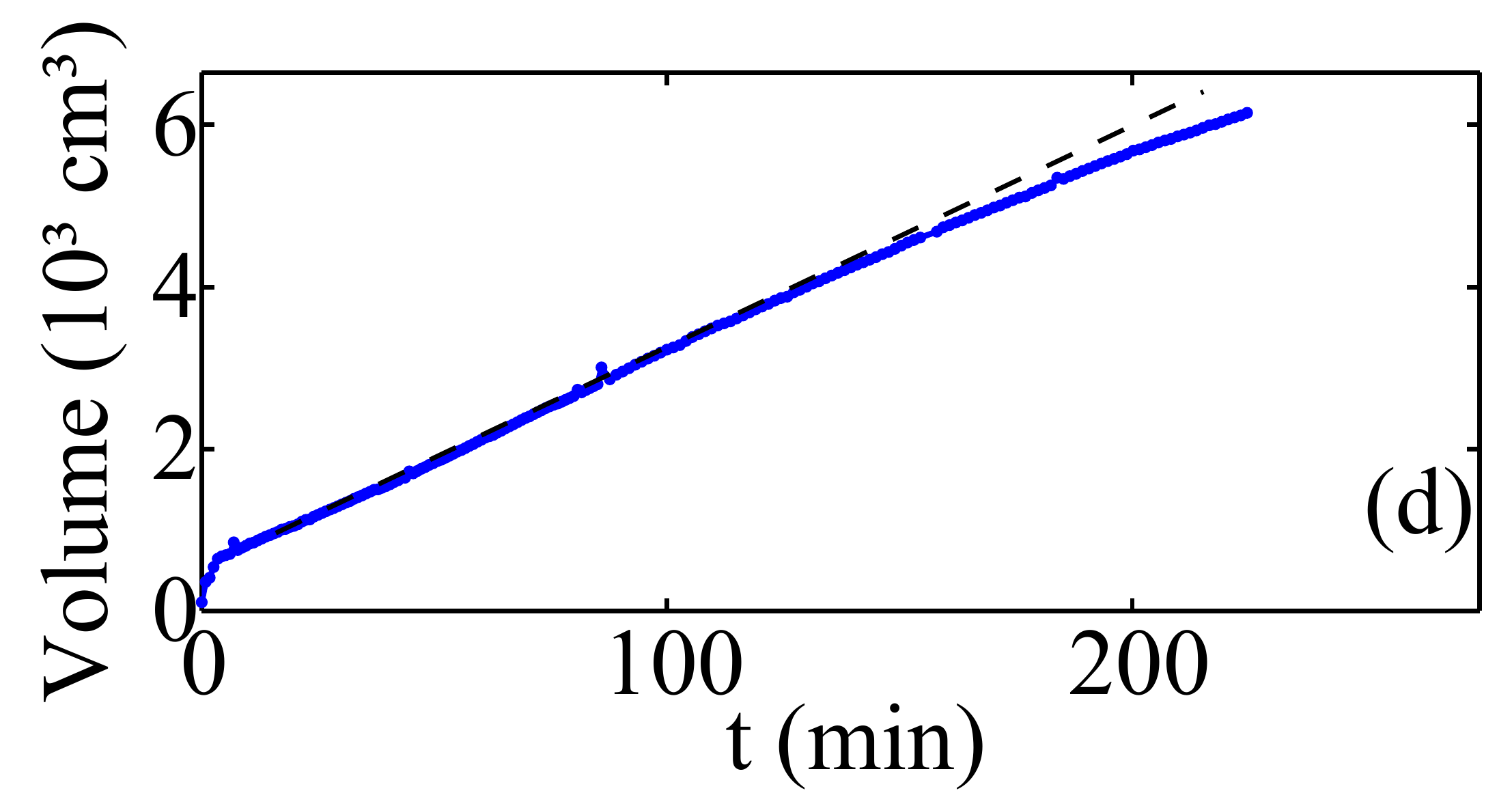}\\
        \caption{(Color online) Comparison of channelization dynamics where $d_{outlet} > d_{cap}$ when the seepage erosion is generated by an upstream flow  and by rain. 
         (a,b) Channel length and eroded volume evolution for $P=0$ with $d_{outlet}=8$\,cm, $\delta_H=6.8$\,cm. 
        (c,d) Channel length and eroded volume evolution for $P = P_1$ with $d_{outlet}=8$\,cm, $\delta_H=2.0$\,cm. 
}
    \label{fig8}
   
    \end{center}
 \end{figure}

To characterize the channel dynamics, we use the channel length $L_{ch}$ defined as the distance from the outlet to a point along the $y$-axis where the depth is above $1$\,cm i.e. $d(0,L_{ch})=1$\,cm. Fig.~\ref{fig6}(a,c) shows the channel length as a function of time for the case with imposed upstream flow and rainfall, respective. One can observe that in case of upstream flow, the graph has an upward curvature indicating that the channel velocity increases as it grows. This can be understood from the fact that the channel tip approaches the boundary where water is injected. Because a significant amount of water flows below the channel surface, the closer the channel is to the source of upstream flow, the greater is the relative importance of surface flow which results in greater erosion. On the other hand, the curve for the rainfall case has a downward curvature. This occurs because a significant fraction of the rain starts to fall inside the channel decreasing the seepage contribution which in turn reduced the growth of the channel. 
Because the choice of the direction is somewhat arbitrary and may be influenced by the splitting of the channel, we also calculate a measure of the scale of the channel using the volume of the grains eroded in the channel as a function of time. The data were obtained by integrating the measured depth field and are plotted for the two cases in Fig.~\ref{fig6}(b,d). The trends shown by the channel length graphs are reflected in these graphs as well. One sees that the velocity of the length grows less slowly compared with the erosion rate, indicating that the channel with rain grows wider with time, indicating a bifurcation.   
 
\subsection{Effect of cohesion on channelization}
\label{deep}

We next investigate how channelization occurs when the channel depth is large compared to capillary rise, and a cohesive layer is present above the water table because of capillary bonds between grains. Due to seepage flow below the unsaturated bed which undermines the bed, the bank becomes mechanically unstable and collapses~\cite{Fox2007}. We first describe the global effect of increasing the depth on seepage erosion before characterizing the nature of the bank avalanches. 

Figure~\ref{fig7}(a) shows an example of a channel formed with seepage flow derived only from the reservoir ($d_{outlet}=8.0$\,cm and $\delta_H=6.8$\,cm). The shape of the channel  remains narrow and it grows towards the back of the bed with a constant width (in Fig.~\ref{fig7}(c)). Thus, the overall shape is similar to that observed with a shallower outlet. However, there are differences in how the erosion front advances. For example, if one compares the contours for $t=157$\,min and $t=255$\,min, the two contours are identical except around $10\,$cm. This local advance of the contour can be observed at other times as well. By contrast, Fig.~\ref{fig7}(b,d) shows examples of channels formed with uniform rainfall on the bed with $d_{outlet}=8.0$\,cm, and $\delta_H=2.0$\,cm. In this case, the channels appear more or less circular like amphitheater-headed valleys observed in nature in agreement with the notion of an isotropic growth in presence of homogeneous rain. However, as the erosion front becomes comparable to the size of the bed, the growth appears more towards the top left and right corners as in the case for shallow channels shown in Fig~\ref{fig5}(c). In contrast with shallow channel discussed earlier, a bifurcation is not observed possibly because the deeper channels draw more water from the reservoir and still continue to grow towards it.

Next, plotting the channel length for these two cases (Fig.~\ref{fig8}(a,c)), we find that the channel length grows in discrete steps with a decreasing amplitude and time interval between successive avalanches when seepage flow is imposed with a reservoir, whereas the avalanches have a larger amplitude and are less frequent in cases with uniform rain. The temporal evolution of volume (Fig.~\ref{fig8} (b,d)) shows similar trends, but without the steps. Although material breaks from the bank of the channel, it falls to the bottom of channel and is then removed by the surface flow. Consequently, although rapid avalanches change the local channel shape, materials leave the system at a lower steady pace due to surface flow. Further, if we compare volume evolution in the two cases shown where flow is imposed from a reservoir (Fig.~\ref{fig6}(b) and Fig.~\ref{fig8}(b)), we notice that the curves have similar shapes. Comparing the cases with rain (Fig.~\ref{fig6}(d) and Fig.~\ref{fig8}(d)), we find that the volume of removed sediment is roughly linear for deeper channels, and therefore the erosion rate is constant during the experiment. Thus, slowing of erosion velocity is not observed and it appears that the reduction of the surface exposed to rain does not influence erosion rates for deeper channels. This observation further suggests that upstream flow gives significant contribution to seepage erosion in case of deeper channels as they approach the reservoir, and suppresses the bifurcation of the channel.

\begin{figure}
 \begin{center}
     \leavevmode
  \includegraphics[width=8cm]{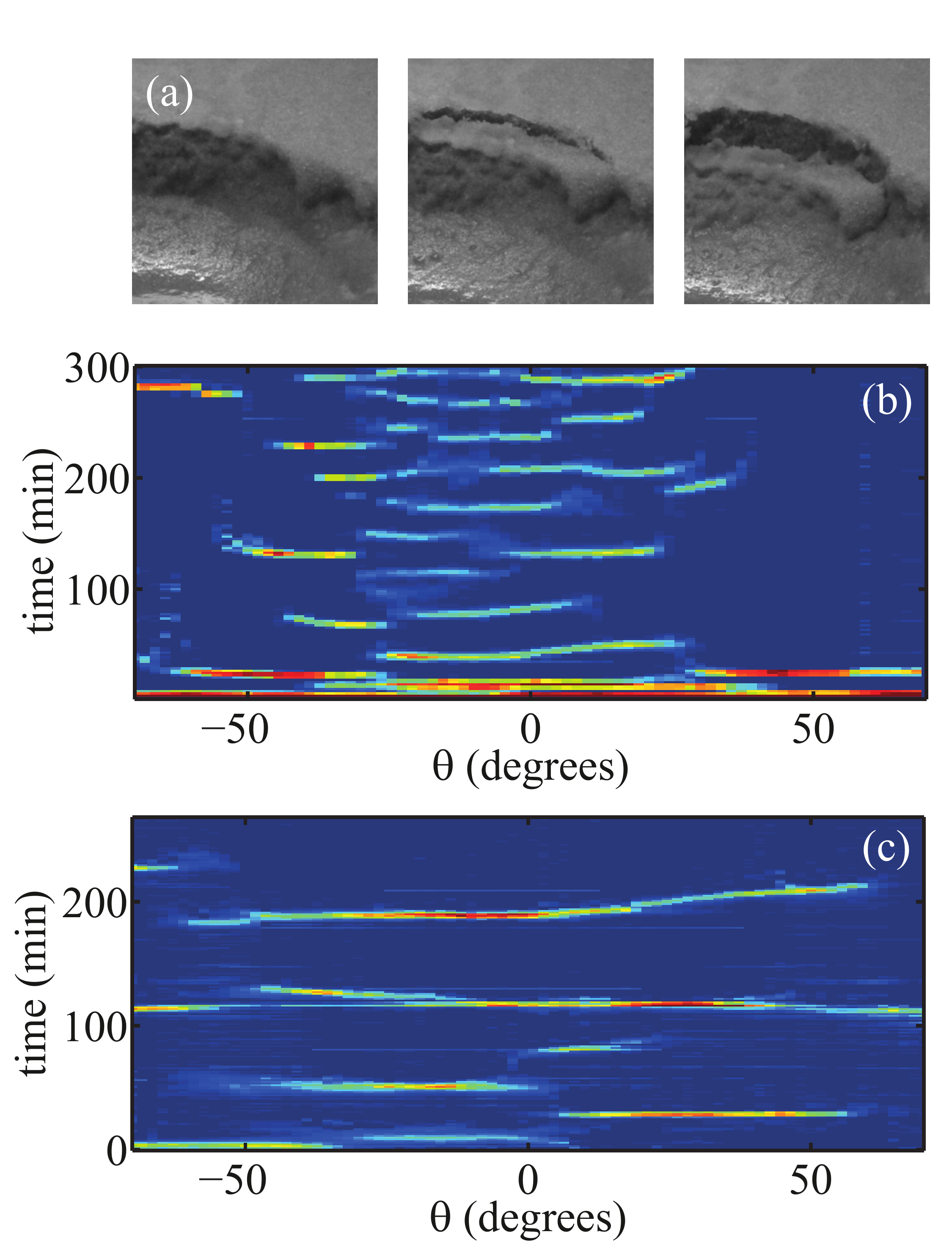}
    \caption{(Color online) (a) Sequence of images illustrate a channel bank collapse. A crack initially appears and breaks a block, which falls inside the channel over the time scale of a minute. (b) The spatio-temporal development of avalanches as a function of the angle $\theta$ and time (b) in case of flow from reservoir and (c) in case with local rain (the experimental parameters are the same as in Fig.~\ref{fig7}). The colorscale is in arbitrary units and corresponds to the temporal derivative of the distance from the contour at $d(x,y)=1$\,cm and origin.
}
    \label{fig9}
    \end{center}
 \end{figure}

An example of a bank collapse in this cohesive regime is shown in Fig.~\ref{fig9}(a) and appears similar to those reported previously~\cite{Fox2007,Threshold}. A block of grains of typical size $10$\,cm - held together by capillary bonds - is observed to break off with a  vertical crack and falls inside the channels. Successive such events produces an irregularly shaped bank with a steep wall as can be observed in Fig.~\ref{fig7}. 
The breaking of a block of material can be understood as follows. 
As the flow in the saturated region of the bed erodes grains, the cohesive strength of the unsaturated region on top (see Fig.~\ref{fig1}) leads the surface to be stable till a threshold is reached. This threshold leads the bank to fail in blocks whose size is given by a balance of the viscous drag below and gravitation which tend to destabilize the bank, and the capillary forces which tend to stabilize it above the angle of repose of the dry grains~\cite{Nowak2005}.   

Because channel length shown in Fig~\ref{fig8}(a,c) detects avalanches only on the central axis, we now use measurement of the full contour defined at an eroded depth $d(x,y)=1$\,cm. For each point of this contour $M$, we compute the distance from origin $d_{M}$ as a function of the angle $\theta$ to the $y$-axis. The time derivative of $d_{M}$ as a function of position and time is plotted in Fig.~\ref{fig9}(b,c). In case of flow from reservoir, avalanches are homogeneously distributed at the beginning and then become increasingly localized in the central direction with decreasing amplitude and increasing frequency. In contrast, avalanches are more regular and of uniform amplitude under homogeneous rain. Collapse events start at the center and then propagate to the edges over a timescale of about ten minutes. 

Although the dynamics are very different in the two cases, we note that the time between two avalanches decreases as the channel grows when the flow comes from a reservoir. In both cases, it appears that the time increases when the amplitude of previous avalanches increases. This is because the channel has to advance further to destabilize the bank after a large avalanche similar to observations in horizontally rotated cylinders~\cite{Caponieri1995}. In the reservoir derived flow, the internal slope inside the channel increases progressively with distance to the origin. Therefore, the vertical size of a block which can fall decreases during channel growth and avalanches become smaller and more frequent. In spite of the intermittent nature of local growth, the overall shape is determined by the distribution of the source of seepage flow over the duration of our experiments. 

 \begin{figure}[h]
 \begin{center}
     \leavevmode
     \includegraphics[width=8.6cm]{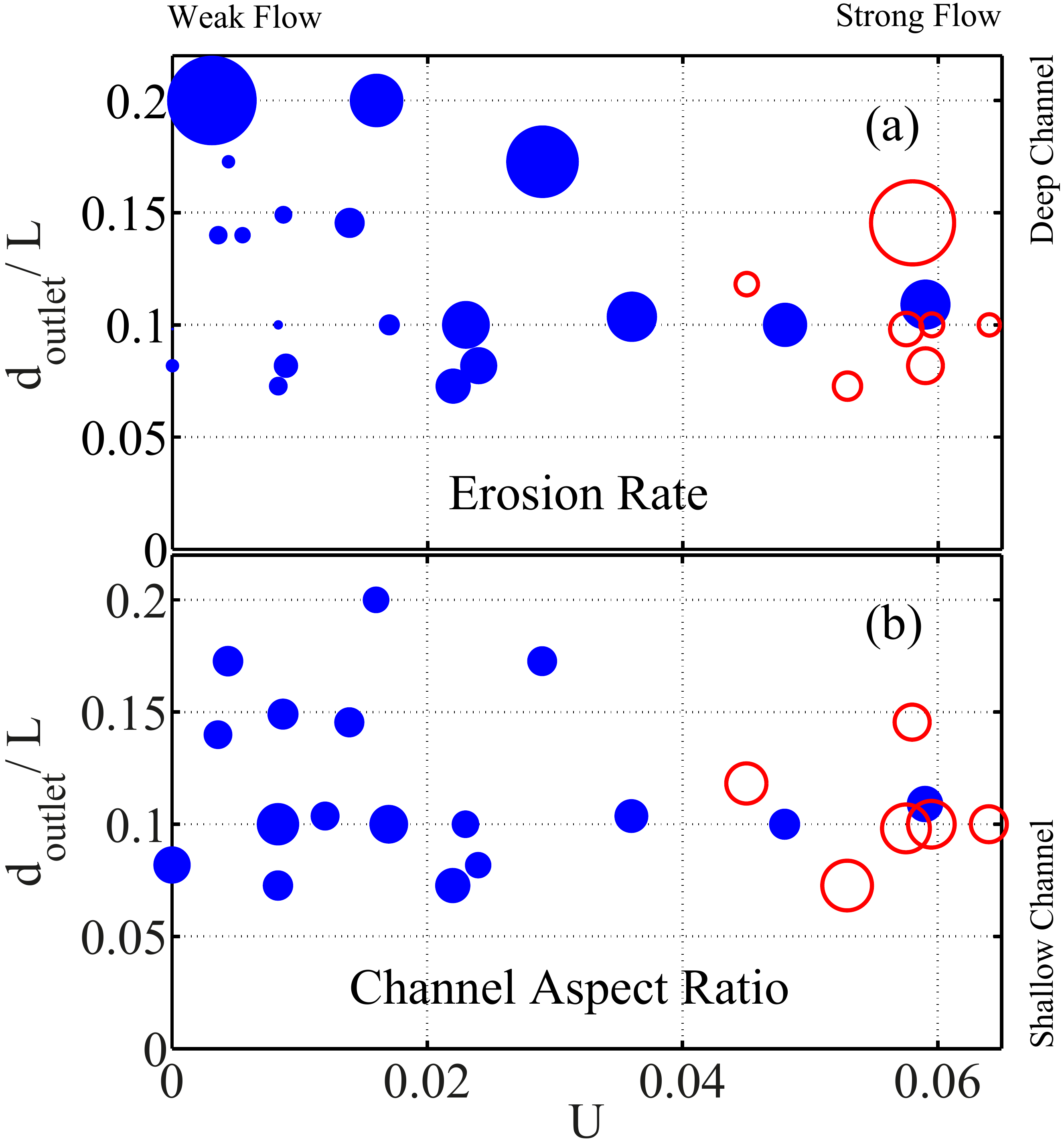}
    \caption{(Color online) (a) Parameter diagram comparing erosion rate $r_s$ of the different experiments processed in case of seepage erosion without rainfall (red hollow circle) and with rain (blue solid circle). The two parameters used are the strength of the upstream flow $U=({H_2}^2-{H_1}^2)/L^2$ and the size of the outlet divided by the length of experiment $d_{outlet}/L$. The size of symbol is proportional to the erosion rate estimated from the volume measurement when the channel length is $20\,$cm. The symbol scale size is ten times greater for measurements with rain. (b) Parameter diagram comparing channel aspect ratio $A_r$ (see text for definition) of the different experiments. The size of symbol is proportional to $A_r$, and the scale is the same for experiments with $P=0$ and $P = P_1$.} 
    \label{fig10}
    \end{center}
 \end{figure} 
 
\section{Analysis of channel shapes}
In order to examine the effect of the experimental control parameters on the erosion rate and the aspect ratio of the channel, we plot these quantities in Fig.~\ref{fig10}. We find that $U$ has to exceed a threshold $\approx 0.04$ for erosion when $P=0$, and erosion occurs for even $U=0$ when $P = P_1$. The erosion rate $r_s$ plotted in Fig.~\ref{fig10}(a) is computed as the time derivative of the channel volume obtained from the topographic maps. In order to compare different experiments, we show $r_s$ corresponding to channel lengths equal to $20$\,cm. For experiments without rain, the average erosion rate  $<r_s>=3.1$\,cm$^{3}$min$^{-1}$ and the corresponding standard deviation is $\sigma_{r_s} =1.9$\,cm$^{3}$min$^{-1}$.
With rain these values become $<r_s>=30$\,cm$^{3}$min$^{-1}$ and $\sigma_{r_s} =20$\,cm$^{3}$min$^{-1}$. Therefore the erosion rate is one order of magnitude higher in the presence of rain for the given rainfall rate. But the high value of the standard deviation shows that other parameters play a role as well. The diagram in Fig.~\ref{fig10}(a) shows in particular that an increase of the upstream flow $U$ in presence of rain augments the erosion rate. Experiments with greater channel depth also give large erosion rates, but in the intermediate regime, no clear trend is visible. 

A channel aspect ratio can be defined as $$A_r=\dfrac{2\,\mathrm{max}(C_y)} {\mathrm{max}(C_x)-\mathrm{min}(C_x)},$$ where, $(C_x,C_y)$ is a parametric representation in the contour $d(x,y)=1$\,cm. With this definition, $A_r=1$ for a semi-circle shaped channel and infinity for a line. Figure~\ref{fig10}(b) displays $A_r$ for different values of experimental parameters when the channel length is equal to $25$\,cm. In order to compare $A_r$ across the various experiments, we evaluate it when channel length is equal to $25$\,cm and find that without local rain, the mean aspect ratio is $<A_r>=2.04$ and its standard deviation is $\sigma_{A_r}= 0.30$, which means that for this channel length, the channel width is close to half of the length. Similarly, we find the values with rain to be $<A_r>=1.54$ and $\sigma_{A_r}=0.22$. Consequently channels in presence of homogeneous rain become wider, as expected. Although $A_r$ appears scattered in Fig.~\ref{fig10}(b), the cases with $P=0$ appear larger than with $P_1$. Also increasing $U$ appears to increase the aspect ratio, but it remains larger than most experiments without rain. Overall, Fig.~\ref{fig10} appears to show that for $P = P_1$, the morphology of the channels created with a combination of rain and upstream flow are similar in shape to that with only rain. Altogether, these examples suggest that the overall channel evolution is not modified by the presence of intermittent avalanches along the bank present in our experiments.  

\section{Conclusions}
Seepage erosion of a flat granular bed is experimentally studied at the laboratory scale to understand the influence of the source of groundwater flow on the shape of channels. Significant differences are found between the case where the ground water comes primarily through a boundary from a far away source and the case where it is fed by uniform local rain. In the first case, when the ground water flow field should be governed by the Laplace equation, the channels formed are elongated with a roughly constant width and point towards the boundary through which water enters the bed. This shape could arise because the tip of the channel drains more water into the channel compared with the sides and therefore most of the channel growth occurs at channel tip. By contrast, if uniform rain occurs on the granular bed, the seepage flow is expected to be governed now by the Poisson equation and water is brought to the outlet almost uniformly from all directions in the half-plane. Therefore, the channel appears to grow uniformly roughly with a semi-circular shape. However, once the channel grows sufficiently long, local maxima of water flux entering the channel appear symmetrically on either side leading to a channel bifurcation. In addition, any perturbation of this growing front could in principle drain more water and therefore is unstable to a fingering instability in the context of the models using the Dupuit approximation~\cite{abrams2009growth,Petroff}. In case of shallow channels, we find the beginning of a splitting under rain which can be easily interpreted by this mechanism. This supports the notion that a channel network can develop in a homogeneous bed whereby ground water flow splits as the channels grow leading the channels to split in turn. 

Further, we have shown the importance of the channel depth in channelization, which is controlled by the
parameter $d_{outlet}$. The upper part of the bed which is unsaturated becomes cohesive due to capillary bonds between grains and collapse as a solid block as its foundations are sapped by the groundwater flow. But, while the spatio-temporal evolution of the erosion front appears to differ strongly, integrated quantities such as the erosion
rate appear similar.

Our observations have important implications for the interpretation of field data because numerous perturbations due to vegetation, pebbles, sediment inhomogeneity are present in nature that could influence channel dynamics. Perturbation of the erosion front due to random avalanching events are shown to not always lead to bifurcations unless they are supported by underlying changes in ground water flow. This occurs because the seepage flow itself is not dependent on surface details. Therefore, our study shows that the overall shape of the channels is primarily determined by the nature of the groundwater flow aided by the convergence of the flow near the channel tip.  

Our experiments also show significant effect of capillarity on the evolution of the channel bank which is relevant to beach rills and spring banks~\cite{Schorghofer,Otvos1999}. When the depth of the channel is smaller or close to the height of capillary rise, the bed is fully saturated and has a smooth appearance. For deeper channels, the upper part of the bed is unsaturated and cohesive due to capillary bonds between grains. In this case, the bank collapses as a solid block as its foundations are sapped by the groundwater flow. The capillary rise effects in the unsaturated region in our experiments may be also representative of similar effects in much larger sedimentary beds where clay can provide greater cohesion between grains and in channels observed in Navajo sandstones in Colorado~\cite{Lamb,LambThesis}.

\begin{acknowledgments}
We thank Joshua Meyer who did preliminary measurements and Samuel King and Fouad Abdul Ameer for experimental assistance at the end of this project, and Daniel Abrams for stimulating discussions. This work was funded by the Department of Energy grants DE-FG0202ER15367(Clark) and DE-FG0299ER15004 (MIT).

\end{acknowledgments}



\bibliography{rainerosion2012}
\end{document}